\documentclass[preprint,review,12pt]{elsarticle}
\biboptions{sort&compress}

\usepackage{caption}

\usepackage{amssymb}
\usepackage{amsmath}
\usepackage{bm}
\usepackage{dcolumn}% Align table columns on decimal point

\journal{Fusion Engineering and Design}

\begin{document}

\begin{frontmatter}


\title{A Doppler backscattering diagnostic for the EXL-50U spherical tokamak: plasma considerations and preliminary quasioptical design}
\author[label1,label2]{Y. H. M. Liang}
\ead{matthewliangyh@gmail.com}
\author[label1,label3]{V. H. Hall-Chen}
\author[label4]{T. L. Rhodes}
\author[label5]{Y. Wang}
\author[label5]{Y. Zhao}

\affiliation[label1]{organization={Future Energy Acceleration \& Translation (FEAT), Strategic Research \& Translational Thrust (SRTT), A*STAR Research Entities, 1 Fusionopolis Way \#20-10 Connexis North Tower},
             city={Singapore},
             postcode={138632},
             country={Republic of Singapore}}

\affiliation[label2]{organization={Blackett Laboratory, Imperial College London, South Kensington},
             city={London},
             postcode={SW7 2AZ},
             country={United Kingdom}}
\affiliation[label3]{organization={Nanyang Technological University, 21 Nanyang Link, \#04-01 School of Physical and Mathematical Sciences},
             city={Singapore},
             postcode={637371},
             country={Republic of Singapore}}

 \affiliation[label4]{organization={Physics and Astronomy Department, UCLA},
             city={Los Angeles},
             postcode={90098},
             state={California},
             country={USA}}

 \affiliation[label5]{organization={ENN Science \& Technology Development Co., Ltd},
             city={Langfang},
             postcode={065001},
             state={Hebei},
             country={China}}


%% Abstract
\begin{abstract}
The EXL-50U spherical tokamak was built by Energy iNNovation to develop technologies for proton-boron fusion (Liu et al., Phys. Plasmas 2024). In tokamaks, turbulence is the dominant mechanism of heat and particle transport. In this work, we designed a Doppler backscattering (DBS) diagnostic to measure plasma flows and turbulent electron-density fluctuations in the EXL-50U. Using the SCOTTY beam-tracing code (Hall-Chen et al., PPCF 2022), we found that operation across the U-band frequency range (40--60 GHz), together with poloidal steering, is capable of accessing a suitable range of fluctuation locations and wavenumbers. A quasioptical system was developed to meet these requirements under physical constraints, including port window availability and in-vessel space. Using the quasioptical system's predicted beam properties, we calculated the attenuation due to the component of the probe-beam wavevector parallel to the magnetic field (mismatch) and determined the degree of mismatch that can be tolerated. We found that toroidal steering is required to minimise mismatch at the relevant cutoff locations, thereby maximizing the backscattered signal. This alignment is particularly important due to the high magnetic pitch angle of the EXL-50U, $\sim35^{\circ}$ at the outboard midplane. The beam properties were also used to predict the spatial resolution of high-$k$ measurements, which we found to be satisfactory. Hence, the DBS system is capable of measuring scattering locations of $0.15 < \rho < 1$, with corresponding turbulent wavenumbers of 0.24~mm$^{-1}$$<$ $k_{\perp}$ $<$ 0.95~mm$^{-1}$. Here, $\rho$ is the normalised radial coordinate of the scattering location, and $k_{\perp}$ is the measured fluctuation wavenumber. 
\end{abstract}


%%\item Presented design elements of a DBS for the EXL-50U spherical tokamak
%\item Found that toroidal steering is needed for magnetic pitch angle matching
%\item Recommended a U-band quasioptical system that satisfies space constraints.


%% Keywords

\end{frontmatter}



\section{Introduction} \label{sec:Introduction}
When a plasma confined in a tokamak is heated to high temperatures, it is susceptible to anomalous transport of heat, particles, and momentum \cite{Hillesheim_2012, doyle2007plasma, garbet2004physics, garbet2010gyrokinetic}, which are caused by various small-scale instabilities driven by temperature and density gradients \cite{doyle2007plasma,international2012iaea, garbet2010gyrokinetic, garbet2004physics}. This is responsible for degrading the confinement of plasmas, which increases the capital costs of the fusion device \cite{Wade2021-ug}. Hence, it is important to understand turbulent transport in any fusion device.
    
The Energy iNNovation (ENN) XuanLong-50U (EXL-50U) spherical tokamak, an upgraded version of the ENN XuanLong-50 (EXL-50), was built by the ENN Energy Research Institute in China to advance vital technologies for proton-boron fusion in spherical tokamaks \cite{ENNroadmap}, see Table \ref{parameters_tokamak} for its main parameters \cite{SHI_2025}. The EXL-50U tokamak was designed with a small aspect ratio to improve plasma confinement \cite{ENNroadmap}. There are two reasons for this improvement. First, a smaller aspect ratio results in greater toroidicity \cite{Kaye2021-ty, ShiBingren_2003} --- defined by the inverse aspect ratio \cite{ShiBingren_2003} --- which reduces areas with bad curvature in the tokamak \cite{Kaye2021-ty}. Secondly, the smaller aspect ratio in spherical tokamaks results in greater $E\times B$ shearing than in conventional tokamaks \cite{Kaye2021-ty}. These two effects suppress electrostatic drift-wave instabilities at the ion and electron scales \cite{Kaye2021-ty}, thereby improving confinement. However, simulation results predicting ion- and electron-scale transport do not always agree with experimental observations \cite{Howard}. Cross-scale simulations that resolve both ion and electron contributions have shown that ion-scale turbulence can be indirectly amplified by electron-scale turbulence through sub-ion-scale structures \cite{Maeyama, Maeyama_2017}, significantly reducing confinement. In this study, we aim to understand cross-scale turbulent interactions in spherical tokamaks such as the EXL-50U through well-resolved experimental measurements of ion-scale turbulence and, in certain cases, the lower end of electron scale turbulence. Hence, this DBS potentially enables the investigation of cross-scale turbulent interactions.
%Hence, it is crucial to understand cross-scale turbulent interactions in spherical tokamaks such as the EXL-50U through well-resolved experimental measurements of electron- and ion-scale turbulence. 
Furthermore, EXL-50U uses proton-boron fusion, which requires ion temperatures to be much higher than electron temperatures \cite{ENNroadmap}. This regime is expected to yield novel and potentially rich turbulence physics that warrant careful experimental investigation. One such technique is Doppler backscattering (DBS), a diagnostic used to measure turbulence in many tokamaks and stellarators worldwide \cite{Hennequin_ToreSupra, TJ-II, EAST, HL-2A, d-three-d, EAST2, LHD, TCV, MCU, LHD2, Carralero2021, globusM, rhodes:2022:design, shi:2023:mastu, macwan:2024:nstx, conway:2025:plasma}. DBS is able to measure flows \cite{pratt2022comparison} and density fluctuations of intermediate length scales, usually 1 $\lesssim$ $k_{\perp}\rho_{s}$ $\lesssim$ 10, where $\rho_{s}$ is the deuterium ion sound gyroradius, which depends on electron temperature. Furthermore, it can measure turbulent fluctuations in the plasma core, which is challenging. 
\begin{table}
    \begin{center}
    \begin{tabular}{ |c|c| }
    \hline
       \textbf{Parameters} & \textbf{Values}
    \\   
    \hline
        Plasma current & 0.5--1~MA 
    \\   
    \hline
        Major radius & 0.6--0.8~m
    \\   
    \hline
        Toroidal magnetic field ($R$ = 0.6~m) & 1.0--1.2~T
    \\
    \hline
        Aspect ratio & 1.4--1.85
    \\
    \hline
        Elongation & 1.4--2
    \\
    \hline
    \end{tabular}
    \end{center}
    \caption{Main parameters of the EXL-50U spherical tokamak \cite{SHI_2025}.}
    \label{parameters_tokamak}
\end{table}
     
\begin{figure}%% placement specifier
    \centering%% For centre alignment of image.
    \includegraphics[width=13cm]{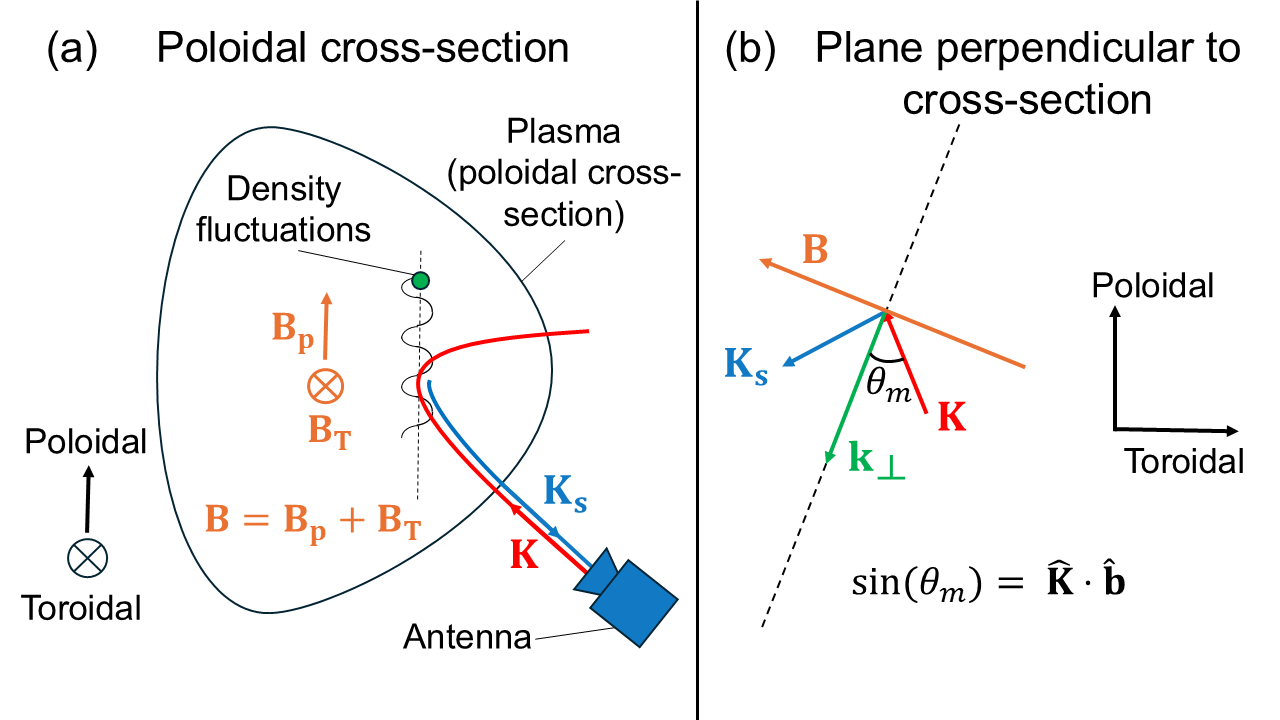}
    
    \caption{Schematic of Doppler backscattering. (a) An antenna emits a microwave probe beam into the plasma and receives the backscattered signal from turbulent density fluctuations. According to the Bragg condition, the turbulence wavenumber is double that of the probe beam's wavenumber, $k_{\perp}$ = $-2K$. Here $k_{\perp}$, $K$ and $K_s$ are the wavenumbers of the turbulence, probe beam, and scattered electric field, respectively, and $\mathbf{B_p}$ and $\mathbf{B_T}$ are poloidal and toroidal components of the magnetic field, respectively. Note that backscattering occurs along path of the probe beam, but as the signal is dominated by scattering from the cutoff location \cite{hallbeamtrace}, in this paper we assume that the backscattered signal only comes from the cutoff. (b) Scattering when the probe beam is not in the plane perpendicular to the magnetic field, that is, if it is mismatched. We define the angle between the normal of the magnetic field and the probe beam's wavevector and the plane perpendicular to the magnetic field to be the mismatch angle, $\theta_m$, such that \textrm{sin$\left(\theta_m\right)$} = $\mathbf{\hat{K}} \cdot \mathbf{\hat{b}}$, where $\mathbf{\hat{K}}$ and $\mathbf{\hat{b}}$ are unit vectors of the probe beam's wavevector and magnetic field $\mathbf{B}$, respectively. 
    }
    \label{fig:DBS figure}
\end{figure}

DBS involves launching a microwave probe beam into the plasma (Fig.~\ref{fig:DBS figure}). The beam is then scattered by electron density fluctuations, and the backscattered electric field is picked up by effectively the same antenna that was used to transmit the microwave beam. The projection of the backscattered electric field onto the antenna's receive function gives the backscattered signal. This backscattered signal is then used to determine the locations of the turbulent fluctuations and their corresponding turbulence wavenumbers ${k}_{\perp}$. We will assume that most of the backscattered signal comes from the nominal cutoff location \cite{Pratt_2024}, which we define as the point where the probe beam wavenumber ${K}$ is minimised. We assume that the measured turbulent fluctuations are located at this point. The exact mechanism of spatial localisation is complicated and beyond the scope of the paper. The measured turbulence wavenumber is determined by the Bragg condition, given by 
\begin{equation}
\label{eqn:Bragg's condition}
    {k_{\perp} = -2K}.
\end{equation}
Here, ${k_{\perp}}$ is the turbulence wavenumber and $K$ is the wavenumber of the probe beam. Hence, to measure a particular turbulence wavenumber ${k_{\perp}}$, a beam with a corresponding wavenumber ${K}$ can be launched, allowing us to receive a measurable backscattered signal from the turbulent fluctuation with that associated wavenumber, ${k_{\perp}}$. Furthermore, varying the poloidal launch angles and frequencies of the beam varies the probe beam's wavenumber at cutoff ${K_{c}}$, allowing us to probe a range of cutoff locations $\rho_{c}$, where $\rho_c$ is the normalised radial coordinate of the cutoff location, and turbulent wavenumbers ${k_{\perp}}$.
In addition to the turbulent density fluctuations, the amplitude of the backscattered signal also depends strongly on the angle between the probe beam's wavevector, $\mathbf{{K}}$, and the plane perpendicular to the magnetic field, $\mathbf{B}$ \cite{Hillesheim_2015, HallChenmismatch, d3d_toroidal, toroidalmismatch}. We refer to this angle the mismatch angle, $\theta_m$, given by,
\begin{equation}
    \label{eqn:Mismatch}
    \textrm{sin$\left(\theta_m\right)$} = \mathbf{\hat{K}} \cdot \mathbf{\hat{b}}.
\end{equation}
Here, $\mathbf{\hat{K}}$ is the unit vector of the wavevector of the probe beam $\mathbf{{K}}$, given by,
\begin{equation}
    \mathbf{\hat{K}}  = \frac{\mathbf{K}}{|\mathbf{K}|},
\end{equation}
and $\mathbf{\hat{b}}$ is the unit vector of the magnetic field $\mathbf{B}$. The backscattered power, $P$, is attenuated when $\theta_m$ is non-zero. The backscattered power is given by an integral over the trajectory of the probe beam \cite{hallbeamtrace},
\begin{equation}
    \label{eqn:power}
    P \propto \int F_i F_m \left| \delta \tilde{n}_e (l) \right|^2 \textrm{d} l.
\end{equation}
Here, $l$ is the arc length along the central ray, $F_m$ is the mismatch attenuation, and $F_i$ is the non-mismatch part of the instrumentation function. $\tilde{n}_e$ is the amplitude of the electron density fluctuations as a function of the backscattered wavenumber, which is given by the Bragg condition at each point along the central ray. This is a simplified version of the backscattered spectral density from \cite{hallbeamtrace}, which is sufficient for this paper. The mismatch attenuation is given by 
\begin{equation}
    \label{eqn:mismatch_atten}
    F_m = \textrm{exp}\left({-2\frac{{\theta_m}^2}{{\Delta \theta_m}^2}}\right), 
\end{equation}
where $\Delta \theta_m$ is the width of the mismatch attenuation; note that $\Delta \theta_m$ depends on the wavenumber, width, and wavefront curvature of the probe beam, as well as the curvature and shear of the equilibrium magnetic field. The beam width is defined as the radius from the beam axis where the electric field drops to 1/e of its on-axis value. The beam curvature is the inverse of the radius of curvature of the beam's wavefront. The explicit formula for $\Delta \theta_m$ is given by,
\begin{equation}
    \Delta \theta_m = \frac{1}{K}\left(\frac{\textrm{Im(}M^{-1}_{yy})}{[\textrm{Im(}M^{-1}_{xy})]^{2} - \textrm{Im(}M^{-1}_{xx})\textrm{Im(}M^{-1}_{yy})}\right)^{\frac{1}{2}}.
\end{equation}
Here $M_{ii}^{-1}$ are the various components of $\bm{M}_w^{-1}$, defined by
\begin{eqnarray}
	\bm{M}_w^{-1}
	=
	\left( \begin{array}{ccc}
		M_{xx}^{-1} & M_{xy}^{-1} & 0 \\
		M_{yx}^{-1} & M_{yy}^{-1} & 0 \\
		0		  & 0		 & 0 \\
	\end{array} \right) \qquad \\
	\qquad  = 
	\left(
	\begin{array}{cc}
		&
		\left(
		\begin{array}{cc}
			M_{xx} & M_{xy} \\
			M_{yx} & M_{yy}
		\end{array}
		\right) ^{-1}
		
		\begin{array}{c}
			0 \\
			0 \\
		\end{array} \\
		
		& \begin{array}{cc}
			\ \ \ \ 0 \ \ \ \ & 0~~
		\end{array}

		\begin{array}{c}
			\ \ \ \ \ \ \ \ \ 0
		\end{array}
		
	\end{array} \right) ,
\end{eqnarray}
where
\begin{equation}
	M_{xx} \simeq \Psi_{xx}
	- K \left( \hat{\mathbf{b}} \cdot \nabla \hat{\mathbf{b}} \cdot \hat{\mathbf{g}} \right)
	,
\end{equation}
\begin{equation}
	M_{xy} \simeq \Psi_{xy}
	- K \frac{\left[ \left( \hat{\mathbf{b}} \times \hat{\mathbf{g}} \right)  \cdot \nabla \hat{\mathbf{b}} \cdot \hat{\mathbf{g}} \right]}{\left| \hat{\mathbf{b}} \times \hat{\mathbf{g}} \right|}
	,
\end{equation}
and
\begin{equation}
	M_{yy} = \Psi_{yy}.
\end{equation}
Here $\hat{\mathbf{g}}$ is the unit vector of the group velocity,
\begin{equation}
	\mathbf{g} = \frac{\textrm{d} \mathbf{q}}{\textrm{d} \tau},
\end{equation}
where $\mathbf{q}$ is the position of the central ray and $\tau$ is a coordinate along that ray. The beam's widths and curvatures are related to the real and imaginary parts of $\bm{\Psi}_w$ and the subscripts indicate their directions, as given in previous work \cite{HallChenmismatch, toroidalmismatch}.

Hence, when the probe beam's wavevector is not perpendicular to the magnetic field vector, the backscattered power is reduced due to mismatch attenuation. A larger mismatch attenuation means that there will be no signal, which is a bigger problem at larger $K$ values \cite{hallbeamtrace}. This is important to note in this work because we want to measure turbulent wavenumbers ${k_{\perp}}$ that are high enough to be considered electron scale. Furthermore, a decrease in backscattered power could either mean a decrease in turbulent fluctuations or an increase in the mismatch angle between the normal to the magnetic field and the wavevector of the probe beam, and these causes are experimentally indistinguishable just from looking at the signal alone. 

In cases where the poloidal magnetic field is significant compared to the toroidal magnetic field, such as in tokamaks with high magnetic pitch angles, the incident beam's wavevector may not be perpendicular to the magnetic field. Hence, understanding mismatch attenuation is important in this work. The EXL-50U tokamak has a high magnetic pitch angle (Fig.~\ref{fig:Bfield}), which means that the mismatch can be large if the toroidal launch angles are not carefully chosen. However, it is possible to use toroidal steering to minimise mismatch for a given beam frequency and poloidal launch angle \cite{d3d_toroidal,toroidalmismatch}. Having steering in both the poloidal and toroidal directions (2D steering) also allows us to potentially make magnetic pitch angle measurements \cite{yeoh2026conceptual}. For $\Delta \theta_m$, the beam width and curvature values are needed to calculate it, and these depend on the design of the DBS system used to propagate a beam of a particular beam width and curvature.

Before designing and installing a DBS system to make these measurements, it is useful to use a synthetic DBS diagnostic to understand how a DBS system would measure turbulent fluctuations in a particular fusion device and to determine the optimal DBS configuration for such measurements. Research has used ray-tracing codes to perform synthetic DBS diagnostics and design DBS systems \cite{Carralero2021}. In ray tracing, the electric field is calculated by propagating a bundle of rays. However, the rays intersect near the cutoff point \cite{Hall-Chen_Parra_Diaz_2021}. The points where the rays meet each other are known as caustics \cite{dodin2025geometrical,lopez2024exact}, and they are problematic because the amplitude of the electric field becomes infinitely large \cite{Hall-Chen_Parra_Diaz_2021} at these points. Consequently, analysis using ray tracing near the cutoff becomes impractical \cite{Honoré_2006}, as DBS requires analysing the properties of the probe beam near the cutoff \cite{MHirsch_2001,EZGusakov_2004}. An alternative is beam tracing, where a single ray is first traced before expanding around it \cite{hallbeamtrace}. This method evolves the trajectory of a Gaussian beam, with the position of the Gaussian envelope's amplitude traced out by the central ray \cite{hallbeamtrace}. The theory of beam tracing has been extensively researched \cite{Casperson, Červený, Kravtsov2007, pereverzev1992use, pereverzev1993paraxial, pereverzev1996, pereverzev1998beam,poli:2001:torbeam, poli:2018:torbeam}, and various phenomena, such as electron cyclotron resonance heating (ECRH) \cite{Prater_2008}, have been simulated using beam tracing. Furthermore, this method can be used near the cutoff \cite{majomar, newruiz}. This paper first conducts synthetic DBS diagnostics on the EXL-50U tokamak with SCOTTY \cite{Hall-Chen:Scotty:2022} in ray-tracing mode to calculate the locations of turbulent fluctuations in the plasma and the corresponding wavevectors that can be accessed. Then, we develop a preliminary quasioptical design for a DBS, while accounting for physical constraints such as the available space for propagating the microwave probe beam. Finally, using beam parameters determined by our preliminary system, we calculate the mismatch attenuation with SCOTTY in beam-tracing mode and determine the optimal toroidal launch angles required. 

The rest of the paper is structured as follows: we present the plasma scenarios used in this study in Section \ref{sec:Plasma scenarios} along with the estimated range of probe-beam frequencies required. We then use SCOTTY, in ray-tracing mode, to calculate the cutoff locations and scattered wavenumbers as a function of poloidal launch angle, see Section \ref{sec:ray_tracing}. We then design the quasioptical system to achieve this range of frequencies and poloidal launch angles, subject to physical constraints; refer to Section \ref{sec:Quasioptical system design}. In Section \ref{sec:beam tracing results}, we use beam-tracing to determine the mismatch attenuation and therefore the toroidal steering required, and also estimate the spatial resolution of high-$k$ measurements. We finally conclude in Section \ref{sec:conclusion}.

\section{Plasma scenarios}
\label{sec:Plasma scenarios}

In this section, we present the EXL-50U plasma scenarios used for our DBS design, with a focus on the density profiles and magnetic equilibrium. We then use these quantities to calculate the cutoff frequencies, estimating the range of DBS frequencies required to measure turbulent fluctuations from the edge to the core.

\begin{figure}
    \centering
    \includegraphics[width=13cm]{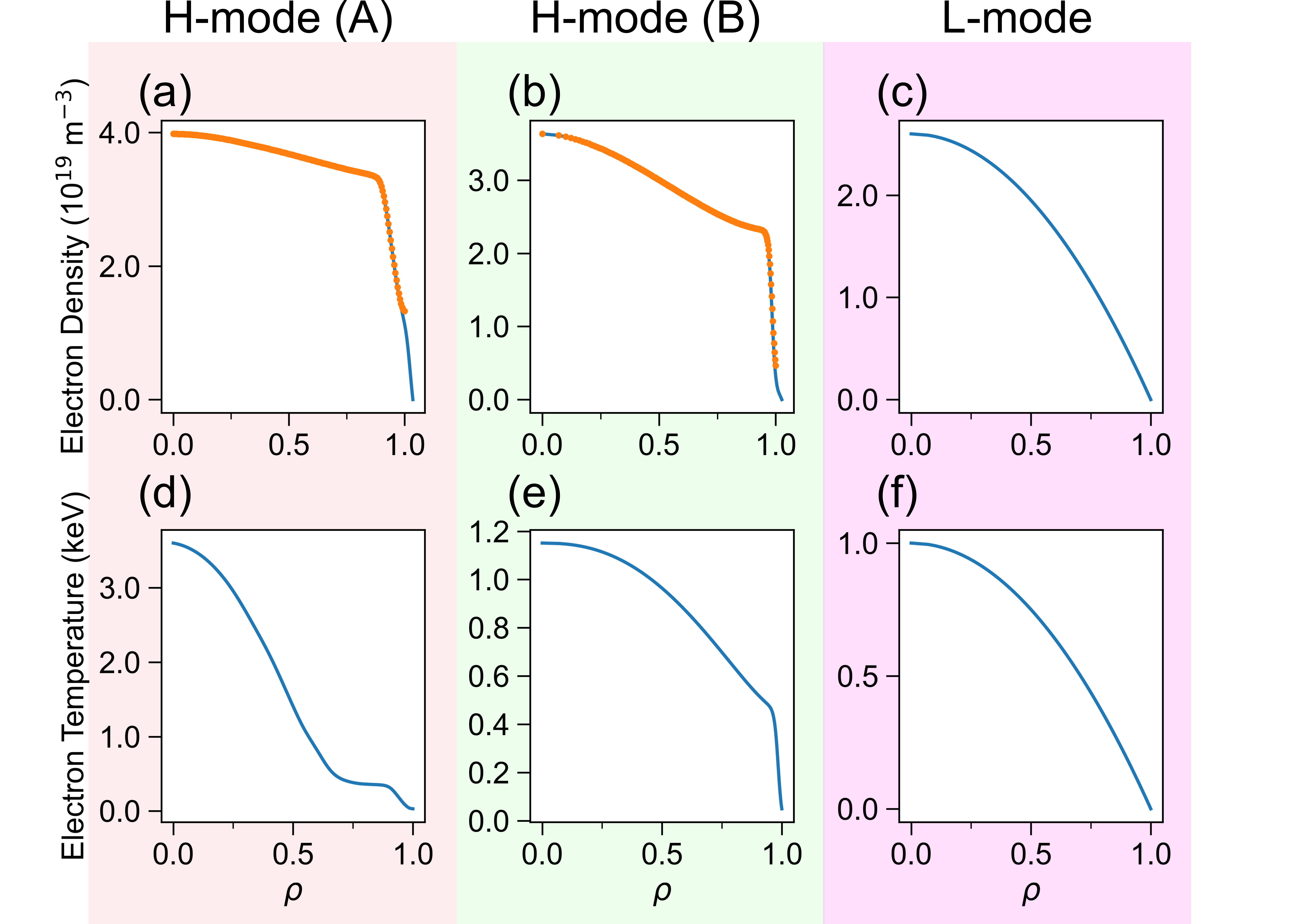}
    \caption{
    Electron density (a)--(c) and temperature (d)--(f) profiles, as functions of the normalised radial coordinate, $\rho$, for the three scenarios used in this paper. These scenarios include the self-consistently modelled H-mode (A), shown in (a) and (d), the lower-density H-mode (B), shown in (b) and (e), and an L-mode, shown in (c) and (f). H-mode (A) has an internal transport barrier due to neutral beam injection. To reach zero density outside the last-closed flux surface, the electron density profiles (orange points) were extrapolated (blue lines) in (a) and (b).
    }
    \label{fig:Profiles}
\end{figure}

The main plasma scenario is a simulated high-confinement mode (H-mode) plasma with neutral beam injection, resulting in an internal transport barrier (ITB) in the temperature profile, as shown in Fig.~\ref{fig:Profiles}(d). We call this scenario H-mode (A). To cover a range of EXL-50U plasmas, we also consider two other scenarios: a H-mode plasma with lower density, called H-mode (B), and an L-mode plasma. The electron density and temperature profiles of these scenarios are shown in Fig.~\ref{fig:Profiles}. All three scenarios use the magnetic equilibrium, given in Fig.~\ref{fig:Bfield}. 

The magnetic equilibrium and profiles for H-mode (A) were self-consistently calculated with integrated modelling; the detailed workflow is shown in Figure 1 of earlier work \cite{Chen_2017}. The following codes were used: EFIT \cite{Lao_1985} for the equilibrium, TGYRO \cite{candy2009tokamak} for the core profile, and NUBEAM \cite{PANKIN2004157}, TORAY \cite{kritz1982ray}, and GENRAY \cite{smirnov1994general} for heating and current drive \cite{Chen_2017}. A modified hyperbolic tangent function was used to estimate the density profile for the lower-density H-mode (B) \cite{Stefanikova:2016:fitting}. Outside the last-closed flux surface, we extrapolated both H-mode density profiles until they reached zero. The L-mode scenario was approximated with quadratic function dependences of electron density and temperature on normalised radial coordinate, $\rho$. The electron density and temperature at the magnetic axis, $\rho = 0$, were chosen to be similar to those of MAST \cite{Hillesheim_2015}, a comparable spherical tokamak, giving $n_{e} = 2.6 \times 10^{19}$~m$^{-3}$ and $T_{e} = 1~\textrm{keV}$.
While we used the same magnetic field profile for all three scenarios as an estimate, we note that strictly speaking, the magnetic equilibrium is only self-consistent with H-mode (A). The magnetic pitch angle is also shown in Fig.~\ref{fig:Bfield}(d); we note that the EXL-50U has a large magnetic pitch angle, $\sim35^\circ$, at the LCFS. This is expected considering that the EXL-50U is a spherical tokamak. The magnetic equilibrium and electron density were inputs for our beam-tracing simulations. On the other hand, the temperature profiles were not used as beam-tracing inputs as relativistic corrections to the electron mass were neglected as $T_e$ is low, lower than 5~keV \cite{H_Bindslev_1992, Wang_Rel}. The temperature profiles were used to calculate the deuterium ion sound gyroradius, $\rho_{s}$, which was then used to calculate the normalised turbulence wavenumber, $k_{\perp}\rho_{s}$, in Section \ref{sec:ray_tracing}.  
\begin{figure}
    \centering
    \includegraphics[width=\linewidth]{figure_3.pdf}
    \caption{Magnetic field profile in the EXL-50U. Here $B_R$, $B_T$, and $B_Z$ are the radial (a), toroidal (b), and $Z$ (c) components of the magnetic field, respectively. The black contour lines show the last closed flux surface and the black crosses in (a)--(c) denote the magnetic axis. The magnetic pitch angle plot on the midplane, $Z=0$, is large and changes significantly with position (d). }
    \label{fig:Bfield}
\end{figure}

\begin{figure}
    \centering
    \includegraphics[width=13cm]{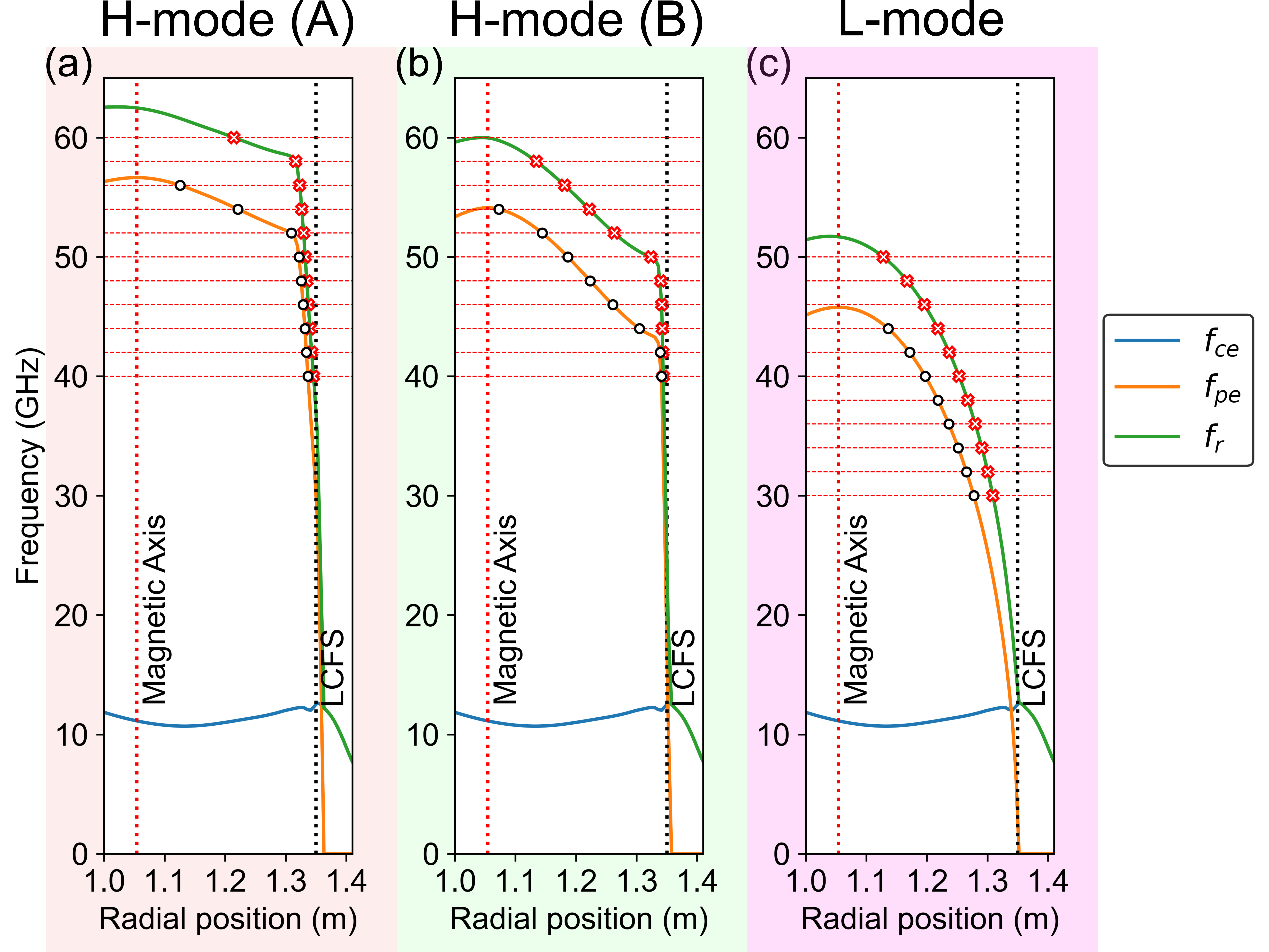}
    \caption{Cutoff frequencies along the midplane in the EXL-50U for (a): H-mode (A), (b): H-mode (B), and (c): L-mode. Here $f_{ce}$ is the fundamental harmonic electron cyclotron frequency, $f_{pe}$ is the plasma frequency, and $f_r$ is the X-mode cutoff frequency. The red dashed lines are the frequencies used, with the black circles and red crosses representing the O-mode cutoffs and the X-mode cutoffs, respectively. In (b), $f_{r}$ is slightly lower than 60~GHz, so a point is not plotted for 60~GHz. A 20~GHz frequency range allows for core to edge coverage. 
	}
    \label{fig:Cutoffs}
\end{figure}

We seek to design a DBS system that can measure density fluctuations from the edge to the core. To that end, we use the magnetic field and electron density to calculate the plasma frequency (O-mode cutoff) and the right-hand cutoff (X-mode) on the midplane for each of the three scenarios, see Fig.~\ref{fig:Cutoffs}. From these cutoff frequencies, we estimate the range of DBS frequencies required, which we then use to calculate the exact cutoff location and measured turbulence wavenumber in Section \ref{sec:ray_tracing}. We find that frequencies between 40~GHz and 60~GHz, the U-band, are suitable for both H-mode profiles; we thus use the U-band at 2~GHz intervals for the beam-tracing simulations. We note that this range allows access to the ITB in H-mode (A) located at around $\rho = 0.60$, where the radial electric field warrants detailed investigation \cite{Hillesheim_2015, Liang:2023:ITB}. For the L-mode profile, frequencies between 30~GHz and 50~GHz can probe turbulent fluctuations at both the edge and the core, so we will also simulate these frequencies at 2~GHz intervals. Since improving H-mode performance is the main goal of this DBS, we later decide, in Section \ref{sec:ray_tracing}, to design for the U-band only, forgoing L-mode edge measurements.

\section{Determining scattering locations and wavenumbers with ray tracing} \label{sec:ray_tracing}
Having downselected the range of DBS frequencies in the previous section, we now use ray tracing to calculate, at each poloidal launch angle and frequency, the:
\begin{itemize}
	\item Scattering location, that is, where the turbulent density fluctuations are measured. This is taken to be at the nominal cutoff, the point along the probe beam's trajectory where its wavenumber is minimised.
	\item Scattering wavenumber, that is, what fluctuation wavenumber is measured at the cutoff, using the Bragg condition.
\end{itemize}
Due to the off-normal incidence of the beams and the launch position being below the midplane, the cutoff locations will not be exactly the same as those shown in the previous section's Fig.~\ref{fig:Cutoffs}. We then finalise the range of frequencies and poloidal launch angles for our proposed DBS. 

\subsection{Ray-tracing simulations} \label{subsec:ray_tracing}      
\begin{figure}
    \centering
    \includegraphics[width=10cm]{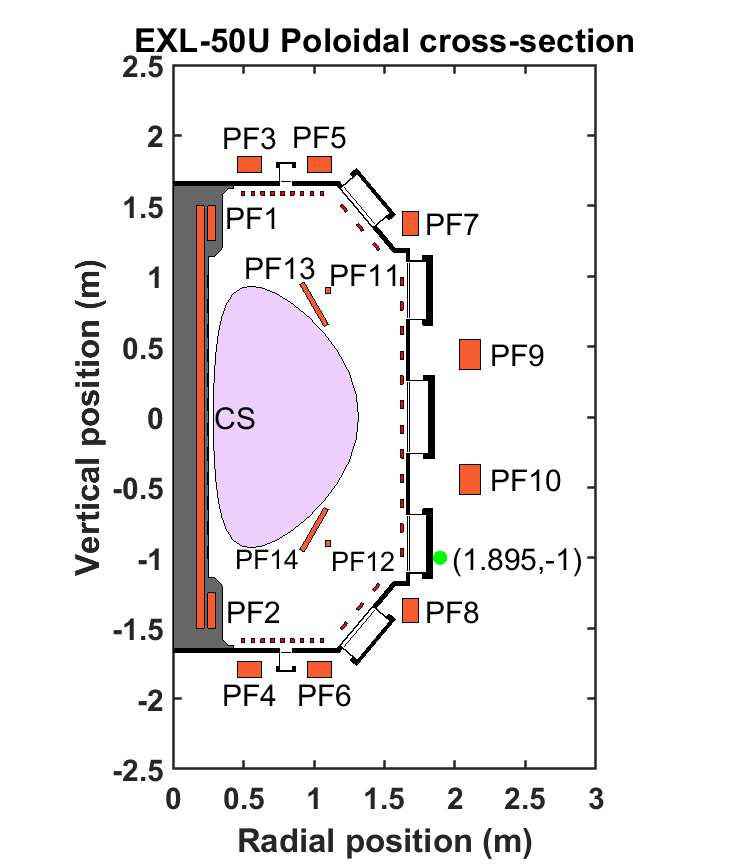}
    \caption{Poloidal section of the EXL-50U containing the plasma (depicted in pink). The launch position is set at $R = 1.9$~m, $Z = -1.0$~m. Microwave probe beam must be launched at appropriate angles to avoid it from being incident on PF14, a coil that goes toroidally around EXL-50U. Note that the plasma shape here differs from that shown in Fig.~\ref{fig:Bfield} as scenario development has yet to be completed. Moreover, the vessel wall is not toroidally symmetric, so we will not consider the vessel wall in our analysis.}
    \label{fig:EXL50Ucrosssection}
\end{figure}

\begin{figure}
    \centering
    \includegraphics[width=13cm]{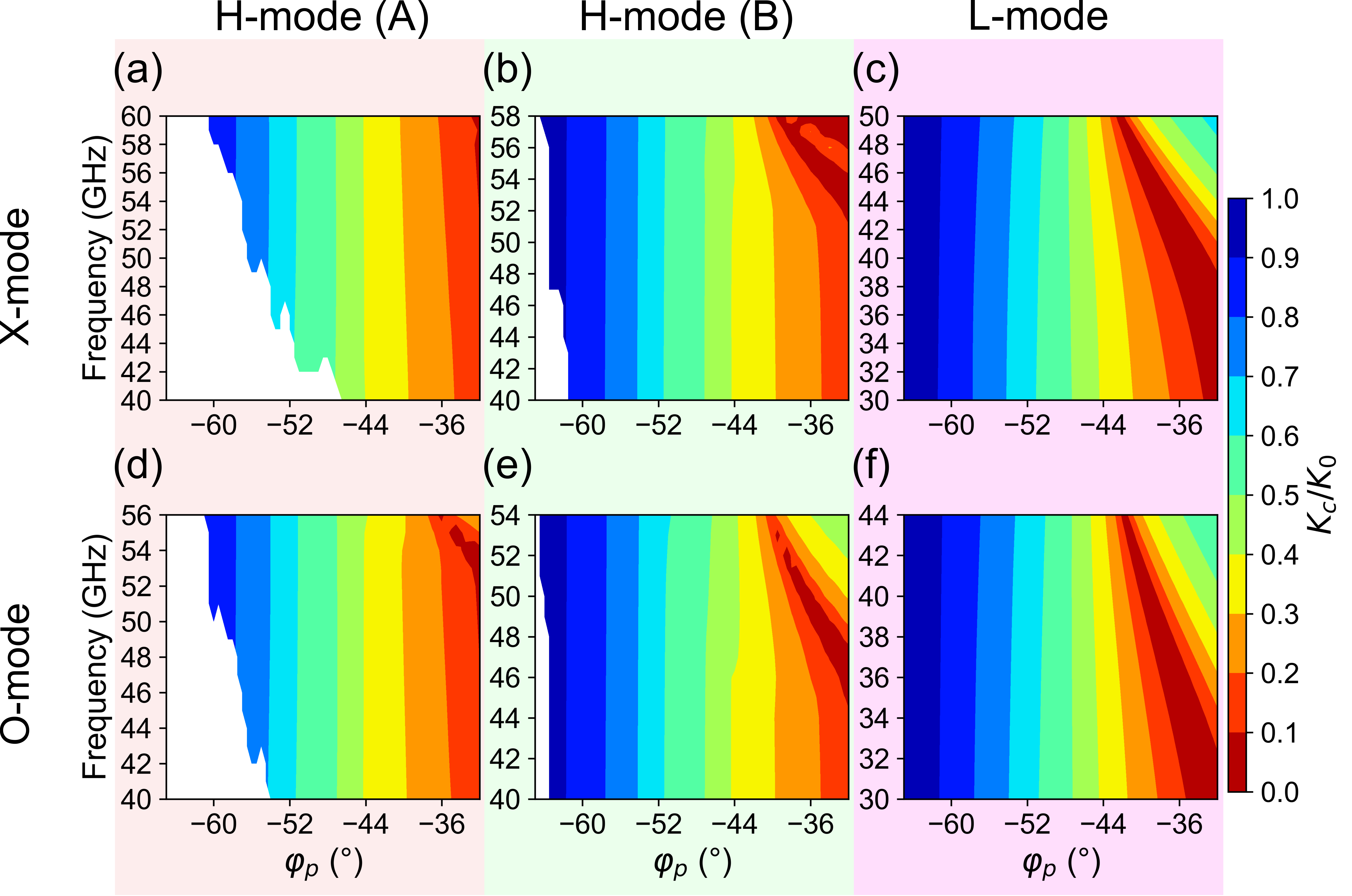}
    \caption{(a)--(c): Ratio of the wavenumber at cutoff to the vacuum wavenumber, $K_c/K_0$, as a function of frequency and poloidal launch angles, $\varphi_{p}$, for X-mode: (a): H-mode (A), (b): H-mode (B), and (c): L-mode. (d)--(f): Same for O-mode: (d): H-mode (A), (e): H-mode (B), and (f): L-mode. The white regions indicate where the cutoffs are located outside the LCFS. We seek to choose poloidal launch angles such that $0.1 \lesssim K_c/K_0 \lesssim 0.4$.
    }
    \label{fig:Kck0}
\end{figure}

We start Scotty ray-tracing simulations at an available below-midplane port, corresponding to $R = 1.895~\textrm{m}$, $Z = -1.0~\textrm{m}$, as shown in Fig.~\ref{fig:EXL50Ucrosssection}. This point also corresponds to the final quasioptical element, which can be the steering mirror, lens, or the antenna itself. In our preliminary quasioptical design, detailed in Section \ref{sec:Quasioptical system design}, the final element is the steering mirror. Nonetheless, the exact quasioptical system is not particularly important for this section, as quasioptics affect the beam width and curvature rather than the ray properties.
For these ray-tracing simulations, the toroidal launch angle is set to zero, $\varphi_{t} = 0^{\circ}$. We now shortlist the range of poloidal launch angles, $\varphi_{p}$. Note that the sign convention for the poloidal angle is defined such that a $0^{\circ}$ poloidal launch angle corresponds to launching the beam parallel to the midplane, and a positive angle corresponds to pointing downwards. When launching from the midplane, increasing the poloidal launch angle increases the wavenumber of the probe beam at cutoff, and thus, by the Bragg condition, increases the measured fluctuation wavenumber. For example, a DBS system on DIII-D has a poloidal steering of up to $20^{\circ}$ \cite{Rhodes:2018:DBSCPS} to measure the fluctuation of high-$k$. At smaller poloidal angles, approximately $\varphi_{p} \lesssim 5^{\circ}$, our DIII-D beam-tracing simulations indicate that there is a significant overlap between the probe beam propagating towards and away from the cutoff, indicating a transition from DBS to conventional reflectometry. Hence, poloidal launch angles that span $5^{\circ}$ to $20^{\circ}$ for a midplane launch are desirable, broadly consistent with typical DBS launch angles \cite{Conway:2025:assessment}. However, our port window is below the midplane. As such, the relationship between the poloidal launch angle and the probed turbulence wavenumber is less obvious. We consider the ratio of the probe beam's wavenumber at cutoff, $K_c$, to its vacuum wavenumber, $K_0$. This ratio, $K_c/K_0$, is the equivalent poloidal launch angle in slab geometry \cite{newruiz}. We seek to have $0.1 \lesssim K_c / K_0 \lesssim 0.4$. The lower bound corresponds to the conventional reflectometry regime and the upper bound to the high-$k$ regime. We calculate this ratio for all three plasma scenarios, see Fig.~\ref{fig:Kck0}.
%DBS relies on strong refraction of the probe beam, increasing the electric field near the cutoff, to attain high spatial resolution \cite{newruiz}.
As such, we determine the poloidal launch angles for each of the three plasma scenarios, given in Table \ref{kck0table}. 
\begin{table}
	\begin{center}
	\begin{tabular}{ |c|c| }
	\hline
		Scenario & $\varphi_{p}$ range
	\\   
	\hline
		H-mode (A) & $-43.5^{\circ} \leq \varphi_{p} \leq -35.5^{\circ}$
	\\   
	\hline
		H-mode (B) & $-43.0^{\circ} \leq \varphi_{p} \leq -40.0^{\circ}$
	\\   
	\hline
		L-mode & $-44.0^{\circ} \leq \varphi_{p} \leq -40.0^{\circ}$
	\\
	\hline
	\end{tabular}
	\end{center}
	\caption{Poloidal launch angle ranges for different profiles. Note that $\varphi_{p} = -28.6^{\circ}$ corresponds to normal incidence on the LCFS.}
	\label{kck0table}
\end{table}
In addition, sufficient clearance should be provided to ensure that no part of the probe beam is incident on the poloidal field coil PF14, as shown in Fig.~\ref{fig:EXL50Ucrosssection}. Based on an empirical guideline used in previous work \cite{Rhodes:2018:DBSCPS}, the beams must be launched such that the minimum distance between PF14 and the beam is at least three times the beam width at the beam trajectory's closest point. Later in section \ref{sec:Quasioptical system design}, we show that it is indeed possible to design a quasioptical system such that this requirement on the beam width is met. Meanwhile, in the rest of this section, we present ray-tracing predictions of the cutoff locations and measured scattering wavenumbers for the range of poloidal launch angles above.

\subsection{Cutoff locations and spectral range} \label{subsec:cutoff_and_spectral}      
\begin{figure}
    \centering
    \includegraphics[width=13cm]{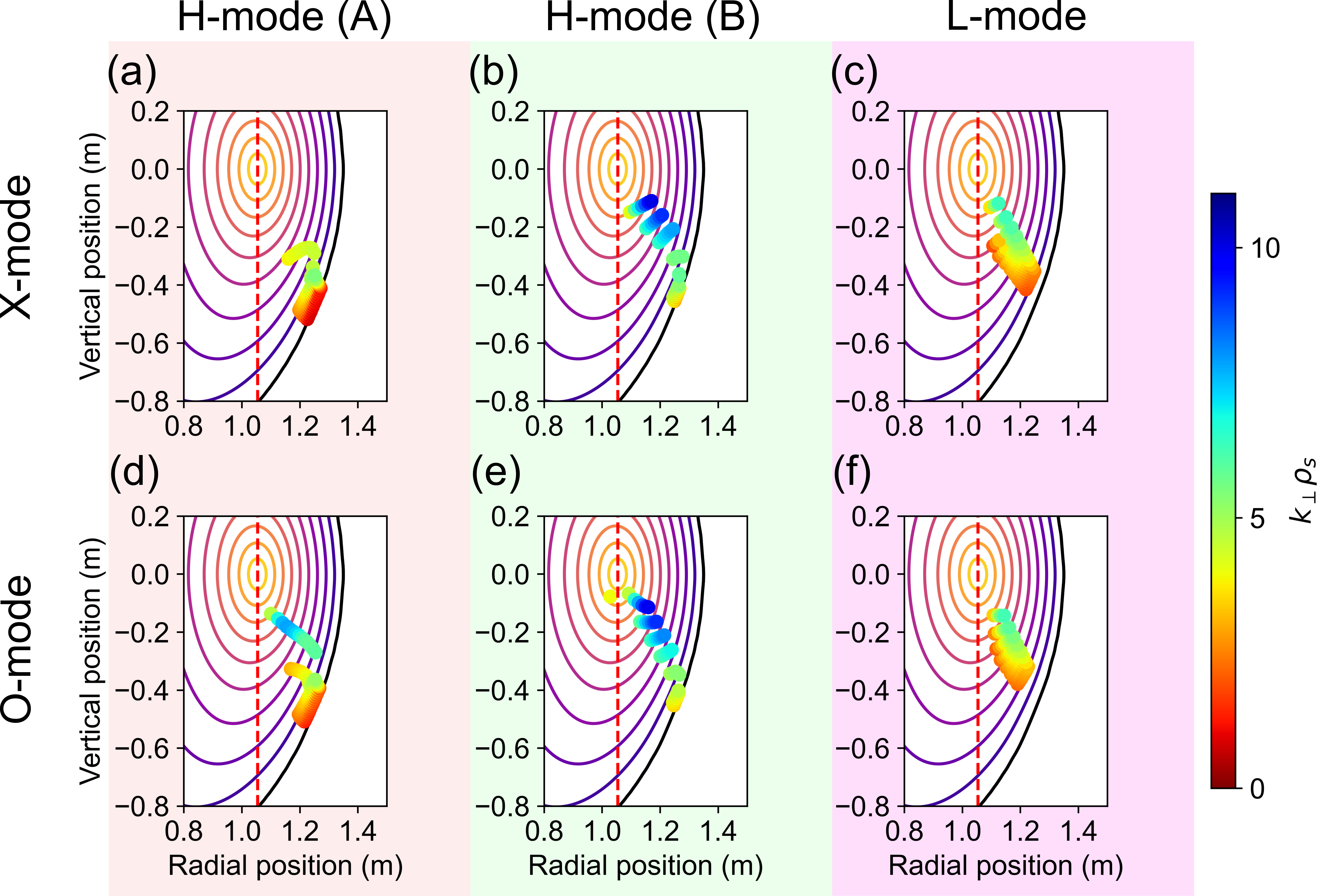}
    \caption{
     (a)--(c): Cut-off positions and their corresponding normalised turbulence wavenumbers $k_{\perp}\rho_{s}$ for X-mode beam trajectories: (a): H-mode (A), (b): H-mode (B), and (c): L-mode. (d)--(f): Same for O-mode: (d): H-mode (A), (e): H-mode (B), and (f): L-mode. The solid black line represents the LCFS ($\rho = 1$), and flux surfaces are plotted in intervals of $\rho = 0.1$. The red dashed line represents the $R$-coordinate of the magnetic axis. 
     }
    \label{fig:spatialrange}
\end{figure}

\begin{figure}
    \centering
    \includegraphics[width=13cm]{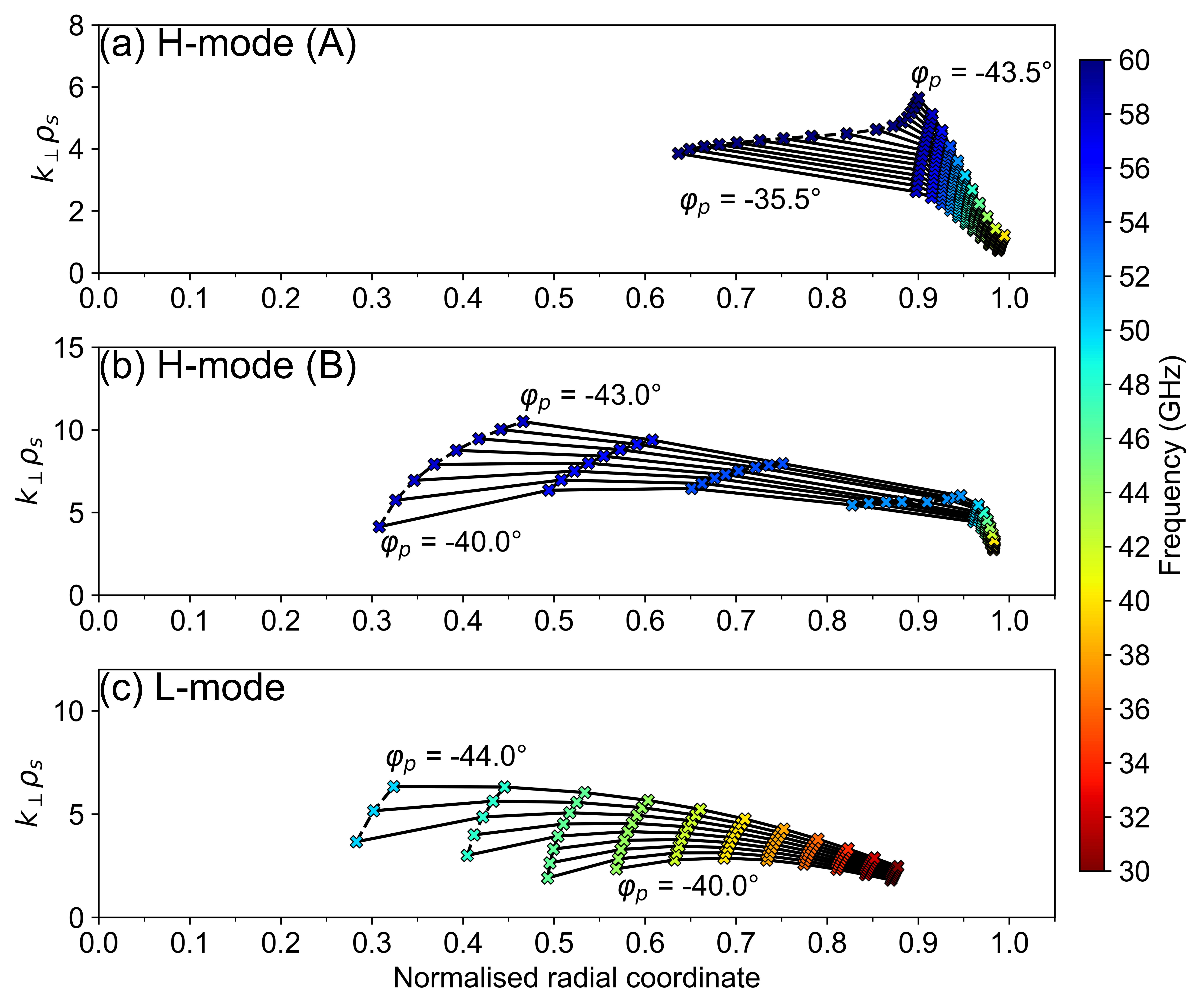}
    \caption{Measured turbulent wavenumbers normalised to the ion sound gyroradius, $k_{\perp}\rho_{s}$, at each cutoff location, as a function of frequency (given by marker fill colours) and poloidal launch angles. Points with the same poloidal launch angles are connected by solid lines, while points with the same frequency are connected by dashed lines. The cutoff locations are given in normalised radial coordinates. All frequencies are in X-mode polarisation, launched at a fixed toroidal angle of 0° for (a): H-mode (A), (b): H-mode (B), and (c): L-mode. 
    }
	\label{fig:polfluxrange_X}
\end{figure}

\begin{figure}
        \centering
        \includegraphics[width=13cm]{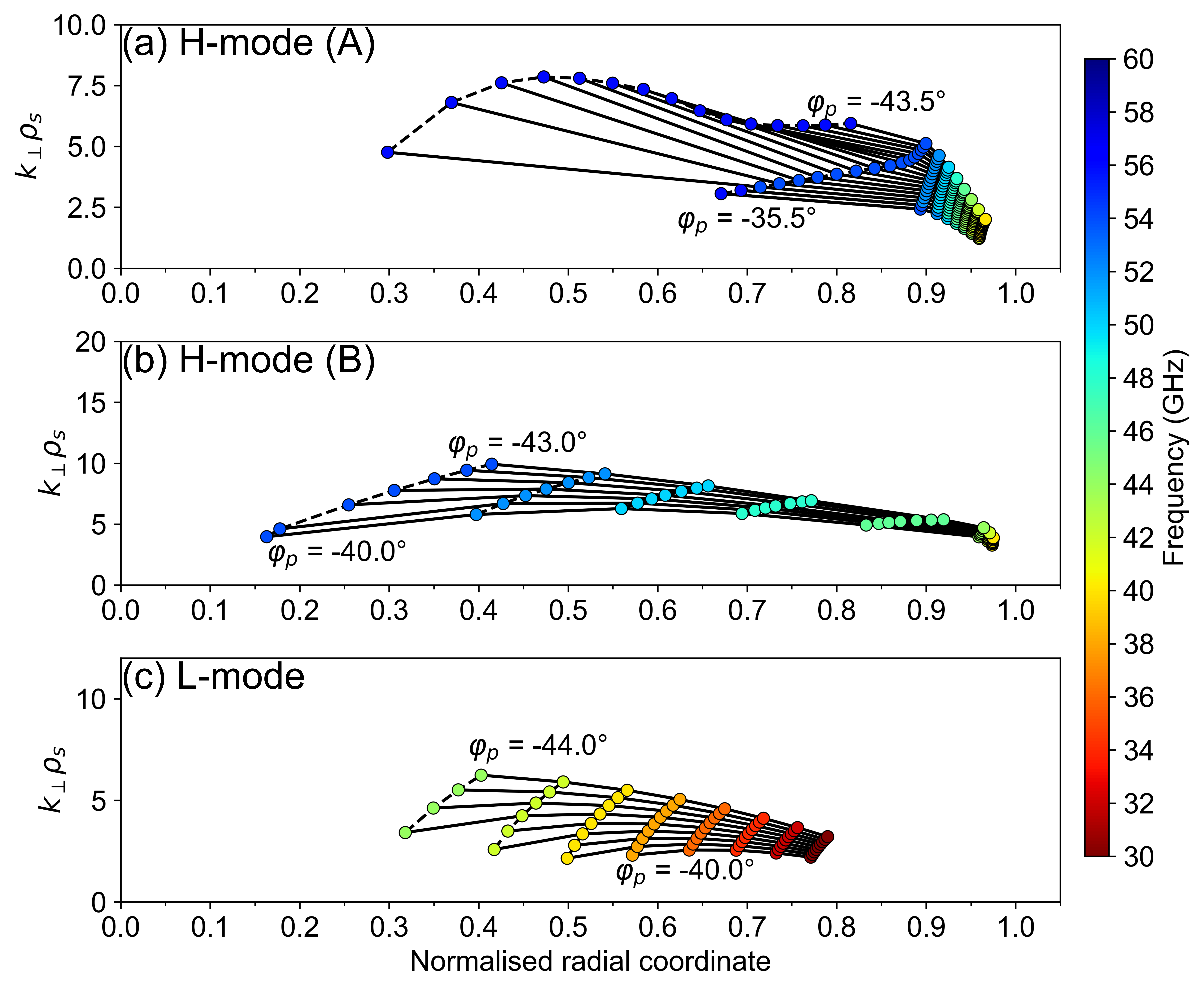}
        \caption{Measured turbulent wavenumbers normalised to the ion sound gyroradius, $k_{\perp}\rho_{s}$, at each cutoff location, as a function of frequency (given by marker fill colours) and poloidal launch angles. Points with the same poloidal launch angles are connected by solid lines, while points with the same frequency are connected by dashed lines. The cutoff locations are given in normalised radial coordinates. All frequencies are in O-mode polarisation, launched at a fixed toroidal angle of 0° for (a): H-mode (A), (b): H-mode (B), and (c): L-mode.
        }
    \label{fig:polfluxrange_O}
\end{figure}

We use ray-tracing simulations to calculate the cutoff locations and measured wavenumbers for both X- and O- mode polarisations, for each frequency and poloidal launch angle. The polarisations are chosen by selecting the appropriate solution of the dispersion relation \cite{hallbeamtrace,stix_1962}. We find that the cutoff locations are all below the midplane, on the outboard side of the plasma, see Fig.~\ref{fig:spatialrange}. A detailed discussion of the cutoff locations for the different plasma scenarios is given later in this subsection.
The measured wavenumbers of the density fluctuations are typically $1 \lesssim k_{\perp}\rho_{s} \lesssim 10$ for all scenarios, see Fig.~\ref{fig:polfluxrange_X} and Fig.~\ref{fig:polfluxrange_O}. 
The wavenumbers in units of inverse length are given in Tables \ref{xmodespectralrange} and \ref{omodespectralrange}. One case where $k_{\perp}\rho_{s} > 10$, corresponding to $k_{\perp} = 0.85 \textrm{ mm}^{-1}$, is when a 58~GHz X-mode beam is launched into the H-mode (B) scenario at $\varphi_p = -43.0^\circ$, with fluctuations located at $\rho = 0.44$. Now, electron scale turbulence has a length scale of $ 10 \lesssim k_{\perp}\rho_{s} \lesssim 30$, which means that the DBS can measure both ion scales and the lower end of electron-scale turbulence. It is important to measure turbulence of these length scales in spherical tokamaks because electron temperature gradients (ETG) drive electron-scale turbulence, which then causes anomalous electron heat transport through cross-scale interactions \cite{10.1063/1.3551701, PhysRevLett.101.075001}. Furthermore, ETGs have been predicted to suppress microtearing modes (MTMs) in spherical tokamaks \cite{Maeyama_Howard_Citrin_Watanabe_Tokuzawa_2024, PhysRevLett.119.195002}, so measuring both ion and electron scale turbulence is important to understand both anomalous electron heat transport and cross-scale turbulent interactions in spherical tokamaks \cite{Speirs:2025:highk}. 
\begin{table}
	\begin{center}
	\begin{tabular}{ |c|c|c| }
	\hline
	   Scenarios & $k_{\perp}$ (mm$^{-1}$)  & $k_{\perp}\rho_{s}$
	\\   
	\hline
	    H-mode (A) & 0.36 $<$ $k_{\perp}$ $<$ 0.95  & 0.72 $<$ $k_{\perp}\rho_{s}$ $<$ 5.65
	\\   
	\hline
	   H-mode (B) & 0.33 $<$ $k_{\perp}$ $<$ 0.91  & 2.75 $<$ $k_{\perp}\rho_{s}$ $<$ 10.50
	\\   
	\hline
		L-mode & 0.19 $<$ $k_{\perp}$ $<$ 0.63 & 1.80 $<$ $k_{\perp}\rho_{s}$ $<$ 6.33
	\\
	\hline
	\end{tabular}
	\end{center}
	\caption{Spectral range for X-mode beams for different plasma scenarios.}
	\label{xmodespectralrange}
\end{table}
\begin{table}
	\begin{center}
	\begin{tabular}{ |c|c|c| }
	\hline
	   Scenarios & $k_{\perp}$ (mm$^{-1}$)  & $k_{\perp}\rho_{s}$
	\\   
	\hline
	   H-mode (A) & 0.24 $<$ $k_{\perp}$ $<$ 0.93  & 1.22 $<$ $k_{\perp}\rho_{s}$ $<$ 7.86
	\\   
	\hline  
	    H-mode (B) & 0.33 $<$ $k_{\perp}$ $<$ 0.83  & 3.28 $<$ $k_{\perp}\rho_{s}$ $<$ 9.93
	\\   
	\hline
		L-mode & 0.21 $<$ $k_{\perp}$ $<$ 0.59 & 2.14 $<$ $k_{\perp}\rho_{s}$ $<$ 6.25
	\\
	\hline
	\end{tabular}
	\end{center}
	\caption{Spectral range for O-mode beams for different plasma scenarios.}
	\label{omodespectralrange}
\end{table}

%%TODO: (matt -> DONE)
\subsubsection{H-mode (A) cutoff locations} \label{subsubsec:H-mode (A)} 

The X-mode and O-mode beams can cover the plasma's pedestal, the ITB region located at around $\rho = 0.6$, and some parts of the core, generally covering a poloidal flux range roughly between $\rho= 0.3$ and $\rho = 1$, see Fig.~\ref{fig:polfluxrange_X}(a) and Fig.~\ref{fig:polfluxrange_O}(a).

Coverage of the plasma core is limited due to two reasons. First, only a small subset of the frequencies and poloidal launch angles we use can probe density fluctuations beyond the plasma pedestal, so the core is probed at only a limited number of locations. Only X-mode beams at 60~GHz and O-mode beams above 52~GHz can penetrate the pedestal and reach cutoff in the plasma core, see Fig.~\ref{fig:Cutoffs}(a). Hence, very few spatial locations within the plasma core can be accessed, limiting its coverage. 

Secondly, we see that the cutoff points beyond the pedestal region for the H-mode (A) profile are slightly more dispersed than those near the edge. Past the pedestal region, $f_{pe}$ and $f_{r}$ change slowly with position (see Fig.~\ref{fig:Cutoffs}(a)), leading to a large spatial separation between the high-frequency cutoff points and further limiting spatial coverage. Hence, higher frequency X-mode beams sampled at smaller frequency intervals can be used to probe more density fluctuations near the core, thereby increasing the spatial and poloidal flux coverage. This can be done using a tunable frequency channel.

\subsubsection{H-mode (B) cutoff locations} \label{subsubsec:H-mode (B)}     Both the X and O mode beams can cover the plasma pedestal and some parts of the core, spanning a poloidal flux range from $\rho$ = 0.15 to $\rho$ = 1. For X-mode beams between 40--50~GHz and O-mode beams between 40--44~GHz, the cutoff locations cluster between $\rho$ = 0.95 and $\rho$ = 1, see Fig.~\ref{fig:polfluxrange_X}(b) and Fig.~\ref{fig:polfluxrange_O}(b) respectively. This is because 40--50~GHz X-mode beams and 40--42~GHz O-mode beams measure density fluctuations in the steep pedestal region (see Fig.~\ref{fig:Cutoffs}(b)), so the cutoffs occur at similar radial coordinates, providing sufficient coverage of the plasma edge.

Although the plasma core is accessed over a wide poloidal flux range, the probed locations are sparse. For X-mode beams above 50~GHz and O-mode beams above 42~GHz, cutoff locations for different beam frequencies at a given poloidal launch angle are more dispersed than at lower frequencies, see Fig.~\ref{fig:spatialrange}(b) and Fig.~\ref{fig:polfluxrange_X}(b) for X-mode, and Fig.~\ref{fig:spatialrange}(e) and Fig.~\ref{fig:polfluxrange_O}(b) for O-mode. When $f_{r}>50~\textrm{GHz}$ and $f_{pe}>42~\textrm{GHz}$, $f_{r}$ and $f_{pe}$ change slowly with position, so a slight change in frequency causes a considerable change in cutoff location (see Fig.~\ref{fig:Cutoffs}(b)). This leads to a large spatial separation between the cutoff points, limiting plasma core coverage. To probe density fluctuations between cutoff points in the core, measurements can be sampled at smaller frequency intervals, increasing spatial and poloidal flux coverage. As in H-mode (A), this can be done using a tunable frequency channel.

\subsubsection{L-mode cutoff locations} \label{subsubsec:L-mode}      
In L-mode, the X-mode and O-mode beams provide good coverage of the plasma core, covering a poloidal flux range roughly between $\rho$ = 0.3 and $\rho$ = 0.9, see Fig.~\ref{fig:polfluxrange_X}(c) and Fig.~\ref{fig:polfluxrange_O}(c), respectively. However, this set of frequencies cannot probe the plasma's edge.  An X-mode beam of frequency 30~GHz does not reach cutoff near the plasma's edge, and any X-mode beam launched at a higher frequency and any O-mode beam with a frequency greater than or equal to 30~GHz will have their cutoffs located deeper in the plasma (see Fig.~\ref{fig:Cutoffs}(c)). To probe fluctuations at the plasma's edge, frequencies below 30 GHz are required. As we later opt to use U-band frequencies only, we did not extend the simulation range to lower frequencies.

\subsection{Summary}
Given the ray-tracing results presented earlier in this section, we propose a DBS diagnostic operating in the U-band, which spans the 40--60 GHz range with a tunable frequency channel. This frequency range enables one to probe the edge to core of H-mode plasmas, supporting ENN's goal of attaining high fusion gain in EXL-50U \cite{SHI_2025}. The U-band also enables access to the core of L-mode plasmas, which could be useful for commissioning the DBS. Unfortunately, the U-band is unable to probe the edge of L-mode plasmas. Nonetheless, L-mode edge measurements are not a priority and thus do not warrant designing a DBS system with a wider frequency range at this stage, which would add cost and complexity. Finally, our proposed system will use both O- and X- mode polarisations, as both provide slightly different spatial coverage in the plasma, and it is not difficult to implement.
%choose a tunable frequency range, choose between O and X mode, different access for each and not difficult to do
% We should probaby select one polarisation?

\section{Designing the quasioptics} \label{sec:Quasioptical system design}        
\begin{figure}
    \centering
    \includegraphics[width=13cm]{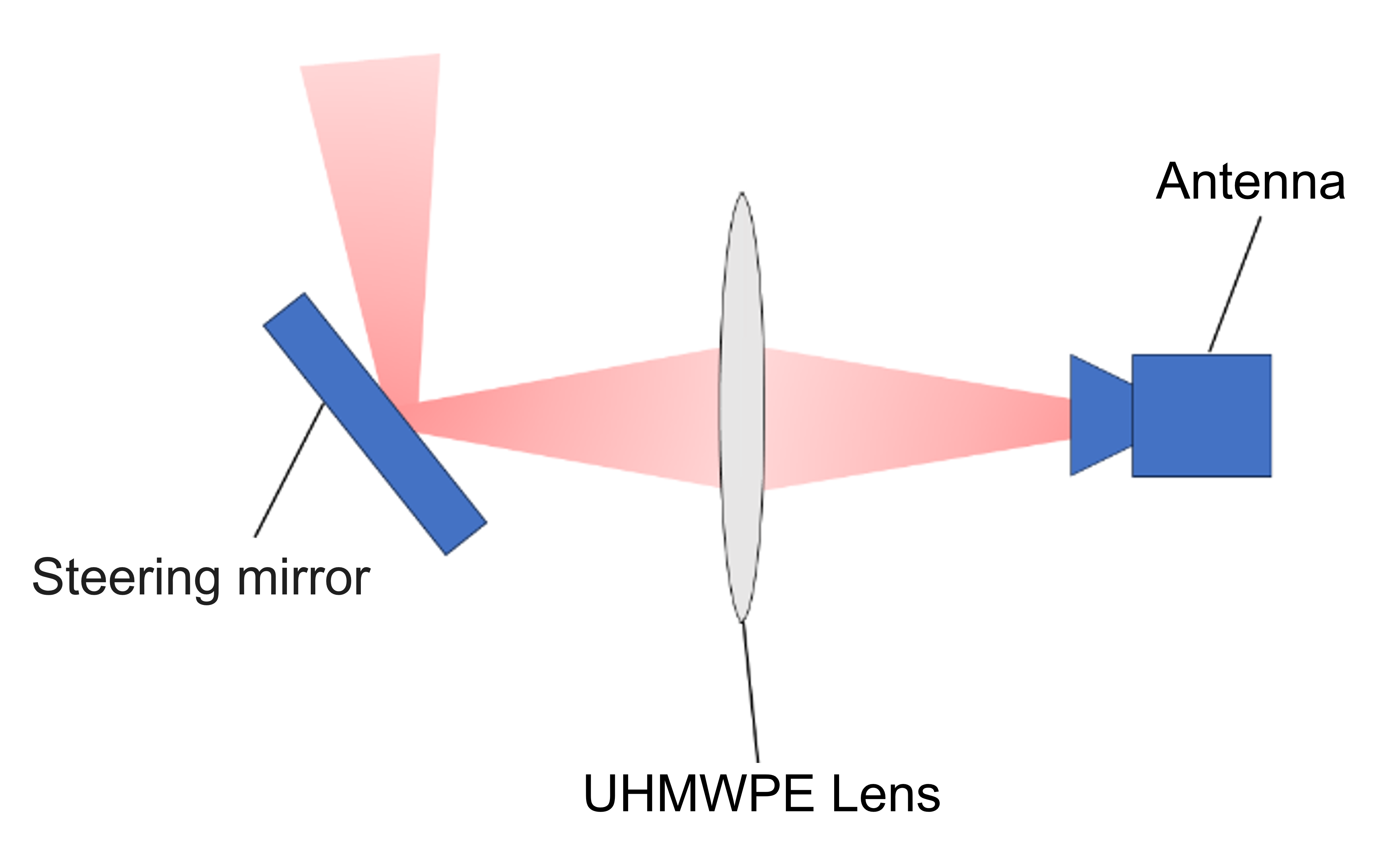}
    \caption{Overview of the DBS used to send microwaves into the plasma. The antenna emits microwaves at various frequencies, which are then focused by a UHMWPE biconvex lens before being reflected into the plasma by a flat mirror with two-axis steering that sets the initial launch angles of the beam. 
	}
    \label{fig:DBS design}
\end{figure}

Having determined the frequencies and poloidal launch angles in the previous section, we now proceed to design the quasioptics for the proposed EXL-50U DBS. This design must satisfy certain constraints, such as having sufficient clearance from the probe beam to propagate. As a starting point, we consider an ex-vessel system consisting of a scalar horn antenna, an ultra high molecular weight polyethylene (UHMWPE) biconvex lens to focus the probe beam and thus reduce its width, and a steering mirror, see Fig.~\ref{fig:DBS design}. This design is similar to DBS diagnostics in the literature \cite{Hillesheim_2015}. 
The rest of this section is divided into three parts. First, we state the design requirements. We then present the detailed quasioptical design, satisfying these requirements. Finally, we discuss the design of the vacuum port window for this DBS.

\subsection{Requirements} \label{subsec:Constraints}      
The EXL-50U DBS has the following requirements:
\begin{enumerate}
	\item At the location where the beam enters the plasma, the beam width cannot exceed 1/3 the distance of that entry point to the top of PF14, otherwise the beam will be incident on PF14. We take the entry point to correspond to the case where the beam is launched at $\varphi_{p} = -35.5^{\circ}$, which corresponds to the lower bound determined for the H-mode (A) profile.
	\item The lens must focus the beam realistically. That is, the minimum beam width should be at least three times its wavelength.
	\item The beam width at the lens must be less than 1/3 the diameter of the lens.
	\item The radius of curvature of the lens must be more than its radius, to enable manufacturability. We will be setting the radius of the lens to be equal to 1.5 times the length of the maximum beam width that will be attained at the lens, since that is the minimum radius the lens can have to not violate the previous constraint.
    \item The projected beam width at the mirror must be less than 1/3 the diameter of the mirror. Hence, we set the mirror radius to be 1.5 times the projected beam width at the mirror.
	\item The diameter of the port window must be three times the beam width to ensure clearance.
\end{enumerate}
The first five requirements will be addressed in subsection \ref{subsec:Proposed DBS system design} and the sixth requirement in subsection \ref{subsec:portwindow}. While not a must, it is desirable for the probe beam to focus at the cutoff location. Unfortunately, this results in the beam being generally wider, making it challenging to fulfil the above requirements. Hence, we will design DBS such that all beams focus at the point where they enter the plasma.

\subsection{Proposed quasioptical system design} \label{subsec:Proposed DBS system design}   
In this subsection, we first explain how we designed the quasioptical system. We then summarise the properties of the proposed system.

To meet the design criteria given in the previous subsection, the following parameters must be chosen wisely:
\begin{enumerate}
	\item Distance from the horn to the lens,
	\item Distance from the lens to the mirror,
	\item Focal length of the lens, and
	\item Aperture radius of the horn.%far-field divergence, given by the full width at half maximum (FWHM) angle.
\end{enumerate}
It turns out that the most difficult requirement to meet is avoiding the beam being incident on the poloidal field coil PF14, requirement 1 in subsection \ref{subsec:Constraints}. Since the beam is closest to PF14 when entering the plasma at $\varphi_{p} = -35.5^{\circ}$, we define the beam to be focused at that point, such that the beam curvature there is zero. Then, the maximum possible beam width is set to 1/3 the distance between the entry point defined in Subsection \ref{subsec:Constraints} and the top of PF14.

With the beam properties defined at the entry, the lens's focal length and the lens-mirror distance constrain the remaining two parameters. To estimate these two parameters, we propagate 40~GHz and 60~GHz beams with the aforementioned beam properties from the plasma entry, back out to the vacuum vessel, through the lens and then to the horn. The focal length and lens-mirror distance are varied to determine the optimal location for the horn, fixing the third parameter: horn-lens distance. The beam widths of the 40~GHz and 60~GHz beams at the horn are then used to determine the horn aperture radius $a$ needed. The 40~GHz and 60~GHz beams correspond to the maximum and minimum beam widths at the horn location, respectively. Hence, we take the mean of the two widths $w_m$ to determine $a$ \cite{goldsmith1998quasioptical}. %The FWHM angle $\theta_{FWHM}$ varies with frequency. Thus, we calculate $\theta_{FWHM}$ for the mid-band frequency of 50 GHz to provide a representative estimate of the FWHM angle across the frequency band \cite{goldsmith1998quasioptical}. 
We obtain
\begin{equation}
	\label{eqn:aperture}
	a = \frac{w_m}{0.644}.
\end{equation}

\begin{figure}
	\centering
	\includegraphics[width=13cm]{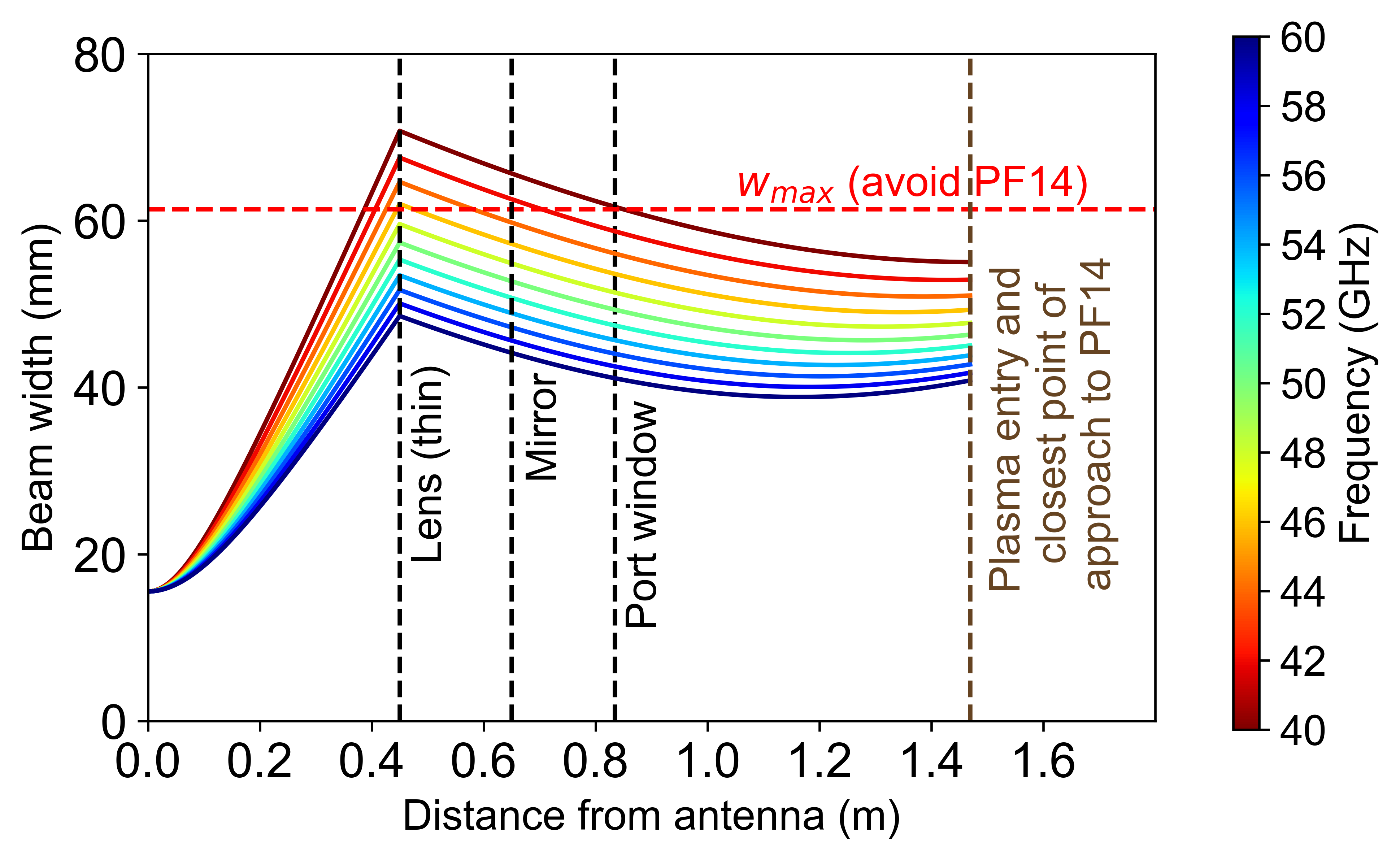}
	\caption{Evolution of the beam width for the 40--60~GHz system as the beam propagates through the quasi-optical system and enters the plasma. Beams are realistically focused, and their widths remain below the maximum value (red dashed line) before incidence on PF14 at the plasma entry (brown dashed line), which is also the closest approach to PF14. This plot shows that we can design a DBS system that meets the physical constraints.
	}
	\label{fig:lenswidth}
\end{figure}

Now that we have the four parameters, we propagate the beam from the horn, through the lens and into the plasma. Next, to check whether requirement 4 can be satisfied, the lens properties must be calculated. As mentioned in requirement 4, the lens radius can be calculated by multiplying the maximum beam width at the lens by 1.5. The radius of curvature of the lens is determined using the thin lens lensmaker’s formula, given by,
\begin{equation}
    \label{eqn:lensmaker's formula}
    \frac{1}{f} =  \left(n-1\right)\left(\frac{1}{R_{1}} - \frac{1}{R_{2}}\right),
\end{equation}
where $f$ is the focal length of the lens, $n$ is the refractive index of UHMWPE, and $R_{1}$ and $R_{2}$ are the radii of curvature of the lens. Note that we have assumed a frequency-independent refractive index of 1.575 for UHMWPE \cite{xie2020quasi}. Since we are using a biconvex lens, $R_1 = -R_2$, so we can determine the radius of curvature of the lens given by,
\begin{equation}
    \label{eqn:simplified lensmaker}
    R_1 = 2\left(n-1\right)f.
\end{equation}
 Now that the lens properties are determined, the four parameters are fine-tuned slightly, as the initial set of values may not fulfil all design requirements. This fine-tuning ensures all design requirements are met and, at the same time, minimise the ratio of the lens radius to $R_1$ to remain well within the thin lens approximation. As a final check, we calculate the evolution of the beam widths of all frequencies, 40--60~\textrm{GHz}, from the horn to the LCFS, see Fig.~\ref{fig:lenswidth}.

\begin{figure}
    \centering
    \includegraphics[width=15cm]{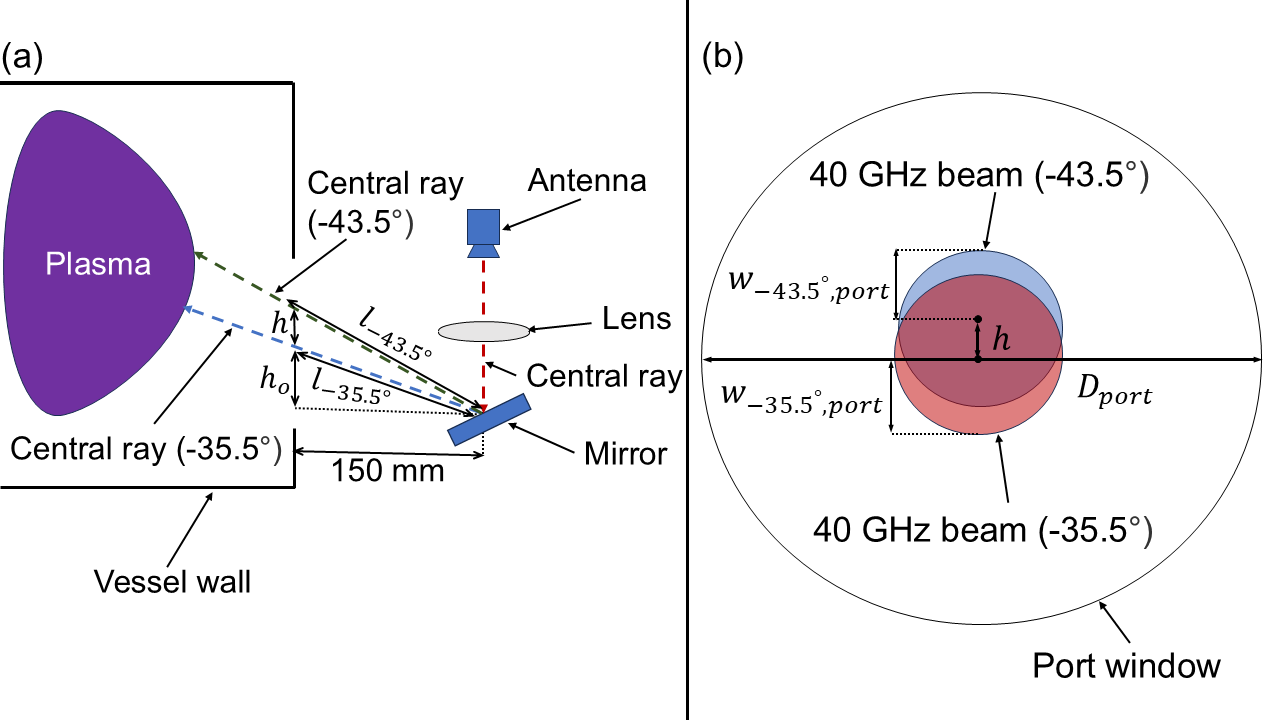}
    \caption{(a) Diagram showing the orientation of the quasioptical system, and the quantities used to calculate the beam widths at the port window and the port window diameter. (b) The port window is large enough to account for not only the beam widths but also the separation between the beams.}
    \label{fig:portwindow}
\end{figure}

Finally, we determine the radius of the mirror. The projected width on the mirror $w_{projected}$ is given by,
\begin{equation}
	\label{eqn:aperture}
	w_{projected} = \frac{w_{mirror}}{\cos{\Phi}}.
\end{equation}
Here, $w_{mirror}$ is the 40~GHz beam width at the mirror, and $\Phi$ is the angle of incidence, defined as the angle between the central ray and the normal of the mirror. $\Phi$ corresponds to the incidence angle when the beam is launched at $\varphi_{p} = -35.5^{\circ}$. To minimise the projected width, $\Phi$ has to be minimised. Hence, the system is oriented such that the antenna launches the beam vertically downward onto the mirror, see Fig.~\ref{fig:portwindow}(a). This configuration yields $\Phi = 27.3^{\circ}$.
The mirror radius can be calculated by multiplying $w_{projected}$ by 1.5. 

Additionally, the far-field divergence of the horn, given by the full width at half maximum (FWHM) angle $\theta_{FWHM}$, can be computed. This is given by \cite{goldsmith1998quasioptical},
\begin{equation}
	\label{eqn:FWHM}
	\theta_{FWHM} = \sqrt{2\ln{2}}\tan^{-1} \left(\frac{c}{\pi w_m f_{beam}} \right).
\end{equation}

Here, $c$ is the speed of light and $f_{beam}$ is the beam frequency. Note that $\theta_{FWHM}$ varies with frequency. Having found the design parameters of our quasioptical components (horn, lens, mirror, and distances between them), we give them in Table \ref{parameters}. 
\begin{table}
\begin{center}
\begin{tabular}{|c|c|}
	\hline
	\textbf{Horn type} & Scalar horn \\   
	\hline
	Aperture radius & 24.1~mm \\   
	\hline
	Beam width at the mouth of horn ($w_m$) & 15.5~mm \\   
	\hline
	Beam curvature at the mouth of horn & 0 m$^{-1}$ \\   
	\hline
	FWHM beamwidth & $6.88^\circ$ -- $10.27^\circ$ \\   
	\hline
	\textbf{Lens type} & Biconvex \\   
	\hline
	Radius of the lens & 106.2~mm \\	
	\hline
	UHMWPE refractive index & 1.575 \\ 
	\hline
	Focal length & 400.0~mm \\   
	\hline
    	Radius of curvature of the lens & 460.0~mm \\
        	\hline

	\textbf{Mirror type} & Planar, 2-axis steering \\ 
    \hline
	Mirror radius & 110.8~mm\\   
	\hline
	Distance between the horn and the lens & 450.0~mm\\   
	\hline
	Distance between lens and mirror & 200.0~mm\\
	\hline
\end{tabular}
\end{center}
\caption{Design parameters for components of the proposed DBS diagnostic.}
\label{parameters}
\end{table}

%biconvex, mirror size

\begin{figure}
    \centering
    \includegraphics[width=13cm]{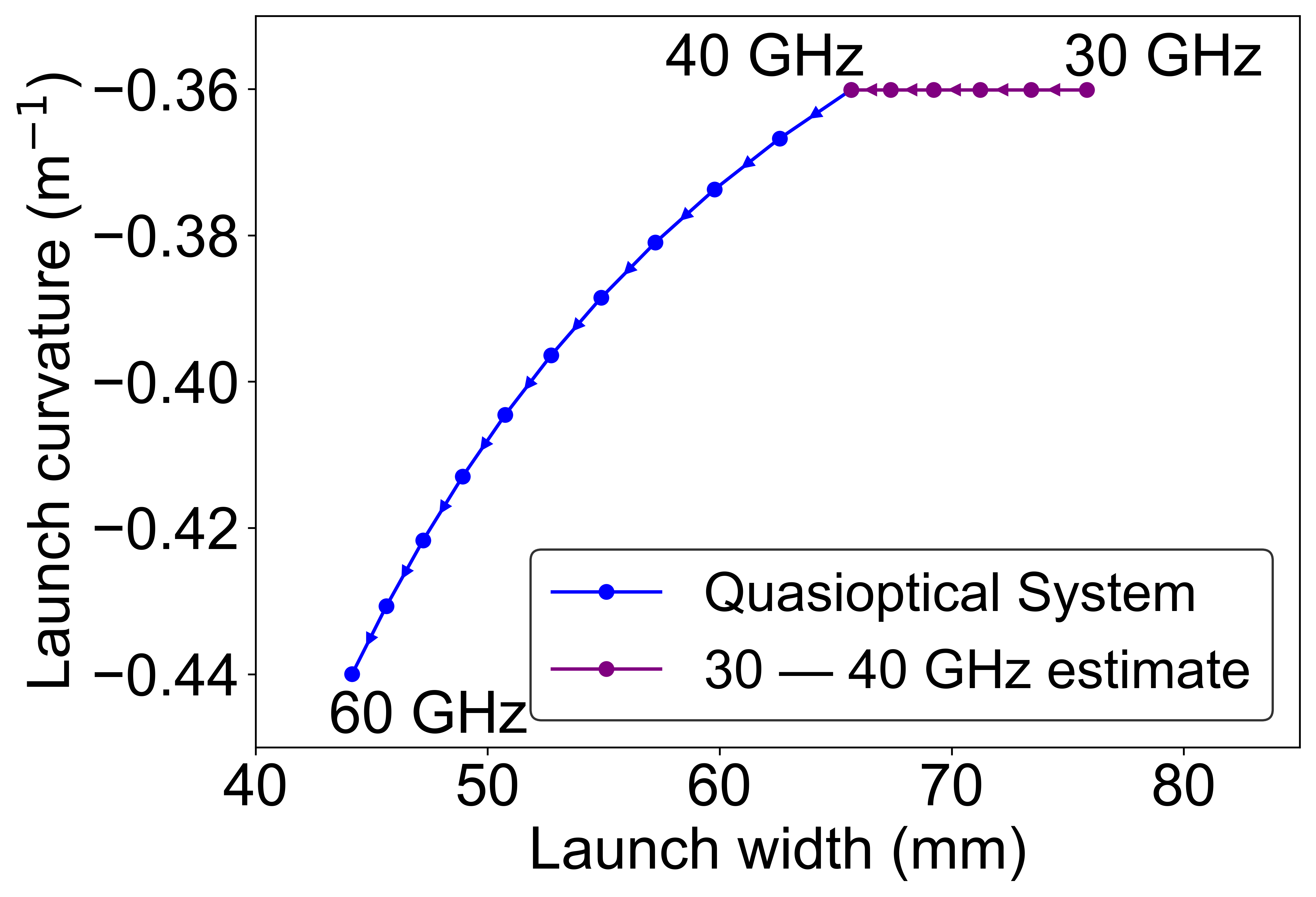}
    \caption{Initial launch widths and curvatures, for each frequency, of the proposed quasioptical system (blue line) and the estimate for low frequencies (purple line). Every frequency has a set of beam width and curvature values, represented by a point, and the frequency varies across points, with arrows indicating the direction of increasing frequency.}
    \label{fig:widthcurv}
\end{figure}

We summarise the launch widths and curvatures for each beam frequency in Fig.~\ref{fig:widthcurv}. 
These beam parameters will be used for SCOTTY simulations in the next section, and will be relevant in subsection \ref{subsec:MismatchA} where we discuss pitch angle matching. For the L-mode profile, our system cannot be used at 30--40~GHz because this is outside the U-band range. Hence, to illustrate pitch angle matching for 30--40~GHz, we use the method below to extrapolate the beam width for frequencies less than 40~GHz. Nonetheless, note that our system is designed for 40--60~GHz. The simulations for 30--40~GHz are solely for illustration purposes and will not be used to evaluate the system's performance.

Beams for 30--40~GHz are launched with an initial beam width $w_0$, given by, 
\begin{equation}
    \label{eqn:launch width}
    w_0 = \sqrt{\frac{z_r\lambda}{\pi}},
\end{equation} 
Here, $z_r$ is the Rayleigh length and $\lambda$ is the vacuum wavelength of the beam. To extrapolate the beam width from 40~GHz, we set $z_r$ such that $w_0$ is equal to the initial beam width of a 40~GHz beam produced by the quasioptical system. This value of $z_r$ is then applied to 30--40~GHz. These parameters were chosen so that the beam does not focus too tightly and thus diverges rapidly from the beam waist. $w_0$ has a frequency dependence that allows for more realistic estimates of the beam widths. We use a converging beam because it improves performance \cite{Bulanin2006}, so the beam curvature is set to $-0.36~ \textrm{m}^{-1}$ for 30--40~GHz, which is equal to the beam curvature of a 40~GHz beam produced by the quasioptical system. These parameters will be used in SCOTTY simulations in the next section.
%w_max,corresponds to case where ray is nearest to the field coil, change the label w_max to avoid field coil

\subsection{Port window design} \label{subsec:portwindow}
As mentioned in subsection \ref{subsec:Constraints}, we will be addressing the final constraint in this subsection, which is that the beam width cannot exceed more than 1/3 the diameter of the port window. The port window must be large enough to account not only for the beam widths but also for the maximum separation between beams when different launch angles are used. We will first compute the beam widths at the port window.

We suppose that the horizontal distance between the launch position and the port window is 150~mm, see Fig.~\ref{fig:portwindow}(a). We also assume that the centre of the port window coincides with the centre of the beam launched at a poloidal angle of $-35.5^{\circ}$. However, the centre of the port window is not at the same vertical height as the centre of the steering mirror. Hence, we need to account for this offset in height $h_o$. We calculate it to be:
\begin{equation}
\begin{aligned}
	h_o &= 150\tan(35.5^{\circ})
\end{aligned}
\end{equation}
Then, the distances travelled by the beams launched at poloidal angles of $-35.5^{\circ}$ and $-43.5^{\circ}$ in millimeters are calculated using:
\begin{equation}
    l_{\varphi_{p}} = \frac{150}{\cos{(\varphi_{p})}},
\end{equation}
where $\varphi_{p}$ is the poloidal launch angle and $l_{\varphi_{p}}$ is the distance travelled by a beam launched at $\varphi_{p}$ in millimeters. Together with the parameters in Table \ref{parameters}, these distances are used to calculate the beam widths for 40~GHz beams at the port window:
\begin{equation}
	w_{-35.5^{\circ},port} = 61.67 \textrm{ mm},
\end{equation}
\begin{equation}
	w_{-43.5^{\circ},port} = 61.24\textrm{ mm}.
\end{equation}
Here, $w_{-35.5^{\circ},port}$ corresponds to the beam width at the port window for a 40~GHz beam launched at $\varphi_{p}$ = $-35.5^\circ$, and $w_{-43.5^{\circ},port}$ corresponds to the beam width at the port window for a 40~GHz beam launched at $\varphi_{p}$ = $-43.5^\circ$. 40~GHz beams are used because this frequency corresponds to the largest beam width, as shown in Fig.~\ref{fig:lenswidth}. Finally, the diameter of the port window $D_{port}$ is calculated by taking the larger of the two beam width values and multiplying it by three:
\begin{equation}
\begin{aligned}
	D_{port}&= 3w_{-35.5^{\circ},port}\\
	&= 185.01 \textrm{ mm}.
\end{aligned}
\end{equation}
We see that a reasonably sized port window can be designed for this system. Now, we verify whether this port window can account for the separation between the beam centres launched at these two different angles $h$ by calculating it:
\begin{equation}
\begin{aligned}
	h &= 150(\tan{(43.5^{\circ})}-\tan{(35.5^{\circ})})\\
    &= 35.35\textrm{ mm}.
\end{aligned}
\end{equation}
$h$ is much smaller than $D_{port}$, so the port window is large enough to allow beams to pass through, see Fig.~\ref{fig:portwindow}(b).

%Now that the preliminary quasioptical design is completed, and with the frequency ranges, poloidal launch angle ranges, and beam parameters determined, we proceed to calculate the scattering locations of the turbulent fluctuations and their associated wavevectors $\mathbf{k_{\perp}}$ that would be measured by this quasioptical system for both O-mode and X-mode, which are chosen by selecting the appropriate solution of the dispersion relation \cite{hallbeamtrace,stix_1962}. These will be discussed in the next section.

%Now that the scattering locations of the turbulent fluctuations and their associated wavevectors $\mathbf{k_{\perp}}$ have been determined, the next subsection will discuss how to find the optimal toroidal launch angle for a given frequency to ensure zero mismatch.

\section{Beam tracing results} \label{sec:beam tracing results}
In this section, we use the initial beam properties from the previous section as input for beam tracing, which we use for two purposes: to calculate the effect of mismatch and to estimate the spatial resolution of high-$k$ measurements.

\subsection{Pitch angle matching and mismatch attenuation} \label{subsec:MismatchA}      
\begin{figure}
    \centering
    \includegraphics[width=13cm]{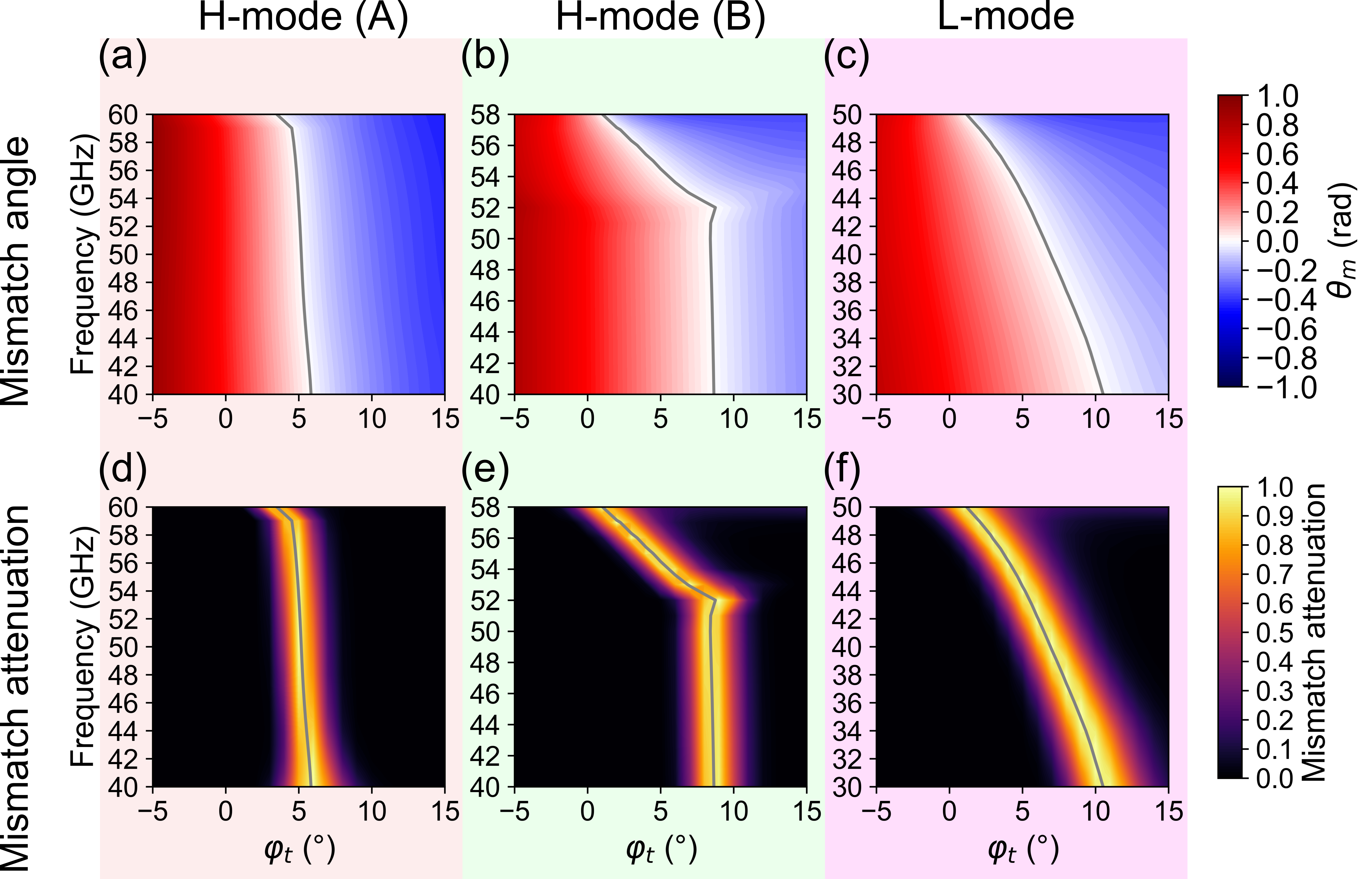}
    \caption{(a)--(c): Heatmaps of the mismatch angle $\theta_m$ of X-mode beams across varying frequencies and toroidal launch angles $\varphi_{t}$: (a): H-mode (A), (b): H-mode (B), and (c): L-mode. (d)--(f): Corresponding mismatch attenuation, defined as $\textrm{exp}\left({-{2{\theta_m}^2}{/{\Delta \theta_m}^2}}\right)$ \cite{hallbeamtrace}: (d): H-mode (A), (e): H-mode (B), and (f): L-mode. The grey line represents zero mismatch. Toroidal steering is needed because it is generally impossible to achieve pitch angle matching for both the core and the edge using the same toroidal launch angle.
    }
    \label{fig:Xmodemismatch}
\end{figure}

\begin{figure}
    \centering
    \includegraphics[width=13cm]{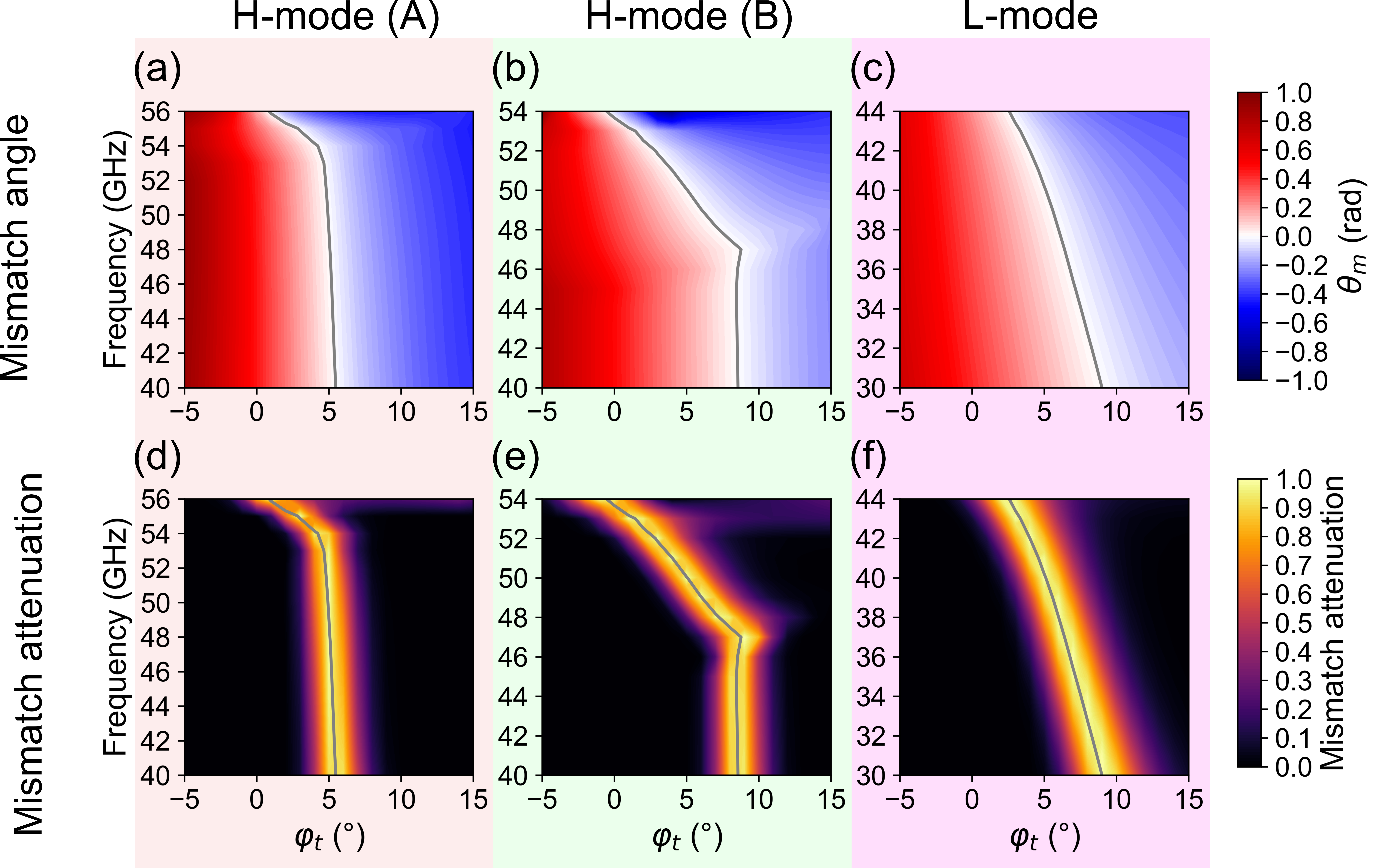}
    \caption{(a)--(c): Heatmaps of the mismatch angle $\theta_m$ of O-mode beams across varying frequencies and toroidal launch angles $\varphi_{t}$: (a): H-mode (A), (b): H-mode (B), and (c): L-mode. (d)--(f): Corresponding mismatch attenuation, defined as $\textrm{exp}\left({-{2{\theta_m}^2}{/{\Delta \theta_m}^2}}\right)$ \cite{hallbeamtrace}: (d): H-mode (A), (e): H-mode (B), and (f): L-mode. The grey line represents zero mismatch. Toroidal steering is needed because it is generally impossible to achieve pitch angle matching for both the core and the edge using the same toroidal launch angle.
	}
    \label{fig:Omodemismatch}
\end{figure}

As the EXL-50U has a large magnetic pitch angle, beams may be severely mismatched if the launch angles are not carefully chosen. Hence, to show the effect of toroidal launch angle on the mismatch attenuation of the beams, we sweep the toroidal launch angles across all frequencies for both X and O modes while keeping other quantities constant. The toroidal response depends strongly on beam properties \cite{d3d_toroidal,toroidalmismatch,Hillesheim_2015}, so the beam parameters at launch must be determined. Furthermore, the dependence of the backscattered power on the mismatch angle follows a Gaussian because there will be a collection of wavevectors within the Gaussian envelope. The implication is that even if the central wavevector does not satisfy the Bragg condition, another wavevector in this Gaussian envelope will. The properties of this Gaussian envelope are thus important, and they depend on the beam width and curvature.
%%%%
Using the beam parameters defined in subsection \ref{subsec:Proposed DBS system design}, beams are propagated at a toroidal launch angle range of $-5^\circ$ $\leq$ $\varphi_{t}$ $\leq$ $15^\circ$, where $\varphi_{t}$ is the toroidal launch angle. We evaluate toroidal steering at three poloidal launch angles: $-36.5^\circ$ for H-mode (A), $-40.0^\circ$ for H-mode (B), and $-43.0^\circ$ for the L-mode. The following subsections discuss how to determine the optimal toroidal launch angle for a given frequency for each plasma profile. Here, the mismatch attenuation defined in Eq. (\ref{eqn:mismatch_atten}) shows how much the backscattered power, expressed as a fraction of the maximum power, is reduced due to mismatch. We calculate the mismatch attenuation for both X- and O- mode beams, see Fig. \ref{fig:Xmodemismatch} and Fig. \ref{fig:Omodemismatch}.

\subsubsection{H-mode (A)} \label{subsubsec:mismatch H-mode (A)}      
For the X-mode beams, the optimal toroidal launch angle does not vary significantly with frequency, so $\varphi_t = 5^\circ$ can be used to avoid severe mismatch, see Fig.~\ref{fig:Xmodemismatch}(a) and Fig.~\ref{fig:Xmodemismatch}(d). The toroidal launch angle for zero mismatch remains nearly constant because the beam reaches cutoff at about the same location in the pedestal for any frequency from 40~GHz to 60~GHz, see Fig.~\ref{fig:Cutoffs}.

However, the optimal toroidal launch angle can generally vary with frequency, as seen in the O-mode case, where it has a noticeable variation in the optimal toroidal launch angle from 54~GHz onwards, see Fig.~\ref{fig:Omodemismatch}(a) and Fig.~\ref{fig:Omodemismatch}(d). This is because when an O-mode beam is propagated with frequencies above 54~GHz, it penetrates the pedestal and reaches cutoff at a very different location, see Fig.~\ref{fig:Cutoffs}. Hence, for a given toroidal launch angle, some frequencies will be matched while others will be mismatched, and the measurable frequency range for each toroidal angle varies and typically does not span the entire 20~GHz range. Therefore, toroidal steering with a tunable frequency channel should be used to prevent severe mismatch.  

In practice, to probe fluctuations in the pedestal, a broad range of beam frequencies (40--60~GHz for X-mode and 40--54~GHz for O-mode) can be used with a toroidal angle of around $5^\circ$ to optimise beams of lower frequencies, since the optimal angle for zero mismatch is the same for these beams. However, to probe a particular location in the core, a narrow frequency range should be used with an appropriate toroidal launch angle. For this specific profile, to probe fluctuations at $\rho$ = 0.30 using O-mode beams launched at $-36.5^\circ$, a frequency range of 55--56~GHz should be used, together with a toroidal launch angle of around $1^\circ$ to optimise these beams. Either way, there is no single compromise angle that provides significant backscattering from both the edge and core channels.  

Finally, we note that the variation of the cutoff location with toroidal launch angle is negligible. Similarly, the measured scattering wavenumber depends only weakly on toroidal launch angle, with 10 to 20\% variation across the studied range. While such variation might be important for interpreting experiments, it does not make a significant difference in the design of our DBS system.

\subsubsection{H-mode (B)}
    \label{subsubsec:mismatch H-mode (B)
}      
The optimal toroidal launch angle for X-mode beams stays nearly constant from 40~GHz to 52~GHz, then decreases from about $8^\circ$ to $1^\circ$ between 52~GHz and 58~GHz, see Fig.~\ref{fig:Xmodemismatch}(b) and Fig.~\ref{fig:Xmodemismatch}(e). This is because X-mode beams from 40~GHz to 52~GHz reaches cutoff at roughly the same location in the pedestal region. Beams above these frequencies reach cutoff at a different location, which explains the large change in toroidal angle that yields zero mismatch. This matches Fig.~\ref{fig:Cutoffs}, where X-mode beams from 40~GHz to 50~GHz reach cutoff at a similar radial coordinate in the pedestal. In comparison, beams with frequencies above 50~GHz can penetrate the pedestal.

The trend is similar for O-mode. The optimal toroidal angle also stays nearly constant from 40~GHz to 48~GHz, but decreases from about $8^\circ$ to $-1^\circ$ between 48~GHz and 54~GHz, see Fig.~\ref{fig:Omodemismatch}(b) and Fig.~\ref{fig:Omodemismatch}(e). Low-frequency O-mode beams also reaches cutoff at a similar radial coordinate, whereas higher frequency beams can reach cutoff at different locations, consistent with Fig.~\ref{fig:Cutoffs}. As in H-mode (A), there is no single compromise angle that yields significant backscattering from both the edge and core channels. Toroidal steering with a tunable frequency channel should be used to prevent beams from becoming severely mismatched.

In practice, to probe turbulent fluctuations in the pedestal region, lower beam frequencies with a broader range (40--52~GHz for X-mode and 40--48~GHz for O-mode) should be used with a toroidal launch angle of about $8^\circ$. To probe a particular location in the core, a narrow range of beam frequencies should be used together with an appropriate toroidal launch angle. For this specific profile, to probe fluctuations at around $\rho$ = 0.30 using a poloidal launch angle of $-40^\circ$ with X-mode beams, a toroidal launch angle of $1^\circ$ should be used together with a narrow frequency range of 56--58~GHz.

\subsubsection{L-mode}
    \label{subsubsec:mismatch L-mode
}      
For the L-mode case, the optimal toroidal launch angle decreases as the frequency increases for both the X-mode and O-mode beams, see Fig.~\ref{fig:Xmodemismatch}(c) and Fig.~\ref{fig:Xmodemismatch}(f) for the X-mode, and Fig.~\ref{fig:Omodemismatch}(c) and Fig.~\ref{fig:Omodemismatch}(f) for the O-mode. This is expected because, from Fig.~\ref{fig:Cutoffs}, whenever a beam with a given polarisation is launched at a different frequency, the beam will always probe density fluctuations at a different location in the plasma. The optimal toroidal launch angle for zero mismatch varies with frequency, so we cannot simultaneously optimise for all locations. This implies that when using the tunable frequency channel to probe a particular location in the plasma, a narrow frequency range should be used with an appropriate toroidal launch angle. For this specific profile, to probe fluctuations located at around $\rho$ = 0.85 using a poloidal launch angle of $-43.0^\circ$ with X-mode beams, a frequency range of 30--34~GHz should be used with a toroidal launch angle of $10^\circ$. However, to probe fluctuations located at $\rho$ = 0.30 using the same poloidal launch angle with X-mode beams, a frequency range of 48--50~GHz should be used together with a toroidal launch angle of around $2^\circ$. 

\subsection{Spatial resolution} \label{subsec:spatial_resolution}
In this subsection, we evaluate the ability of the proposed DBS to measure high-$k$ fluctuations, $k_\perp \rho_s \sim 10$. Such measurements are difficult because mismatch attenuation is higher at high-$k$ \cite{hallbeamtrace, toroidalmismatch}, requiring toroidal steering. Having addressed toroidal steering in the previous subsection, we now consider the other challenge of high-$k$ measurements, spatial resolution \cite{Hillesheim_2015}. We consider the case where a 58 GHz X-mode beam is launched into the H-mode (B) scenario at $\varphi_{p} = -43.0^\circ$ and $\varphi_{t} = 6.0^\circ$. Here, this toroidal angle is chosen to achieve zero mismatch. These parameters correspond to $k_\perp = 0.95 \textrm{ mm}^{-1}$ and $k_\perp \rho_s = 11.0$. 

\begin{figure}
	\centering
	\includegraphics[width=13cm]{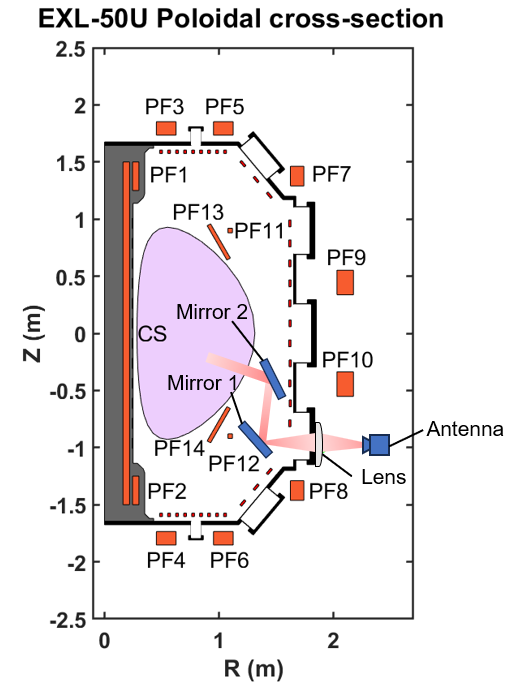}
	\caption{Spatial resolution results for a 58 GHz X-mode beam launched into H-mode (B) at $\varphi_{p} = -43.0^\circ$ and $\varphi_{t} = 6.0^\circ$. (a) Plot of the DBS filter function with distance along the beam path, with the interquartile range of the backscattered power highlighted in orange. 50\% of the backscattered power comes from at most 0.075 m away from the cutoff (red dashed line).
    (b) Beam trajectory in the poloidal plane, with the segment of the beam path corresponding to 50\% of the power highlighted in red. Most of the signal expected for high-$k_\perp$ measurements come from roughly the same radial location as the cutoff.
    %We calculate the interquartile range of the backscattered power, as a function of distance along the beam (b), and find that most of the signal expected for high-k measurements come from approximately the same radial location as the cutoff (b). % Bad phrasing, I should rewrite. Should also include the launch frequency, tor and pol angles here.
	}
	\label{fig:spatial_resolution}
\end{figure}

Next, we plot the DBS filter function along the beam path, which corresponds to $F_i F_m $ in Eq. (\ref{eqn:power}). Then, the interquartile range of the backscattered power, defined as the range between the 25th and 75th percentiles of the integrated DBS filter function, is calculated. Our analysis shows that 50\% of the signal expected for the high-$k$ measurement originates near the cutoff, see Fig.~\ref{fig:spatial_resolution}(a). By determining the segment of the beam path corresponding to the interquartile range of the backscattered power calculated, we find that 50\% of the signal is received from roughly the same radial location as the cutoff at $\rho =0.47$, with the backscattered power coming from at most $\Delta\rho = 0.06$ away from the cutoff location (see Fig.~\ref{fig:spatial_resolution}(b)). This shows that the spatial resolution for this high-$k$ measurement is satisfactory. Hence, we find that the spatial resolution of high-$k$ measurements is indeed adequate.

\section{Conclusion} \label{sec:conclusion}
We completed a quasioptical design for a DBS for the EXL-50U spherical tokamak. Using SCOTTY simulations, we show that the DBS can access turbulent fluctuations in H-mode (A) over a wide range of scattering locations, $0.30 < \rho < 1$, thereby providing access to the pedestal, the ITB region, and parts of the core. It is also able to measure $0.24 \textrm{ mm}^{-1} < k_{\perp} < 0.95\textrm{ mm}^{-1}$. We also show that poloidal flux coverage of the H-mode (B) and L-mode profiles is adequate. Across all profiles, the DBS can measure turbulence on the scale of $k_{\perp}\rho_{s}$ $\lesssim$ 10 with a maximum range of $0.72 < k_{\perp}\rho_{s} < 10.50$. This makes it suitable for measuring ion-scale turbulence and, in certain cases, the lower end of electron-scale turbulence (10 $\lesssim$ $k_{\perp}\rho_{s}$ $\lesssim$ 30), which might potentially enable investigation of cross-scale turbulent transport, which is expected to be important in spherical tokamaks. To probe these turbulent fluctuations, the DBS must operate in the U-band frequency range and a poloidal launch angle range of $-43.5^\circ$ $\leq$ $\varphi_{p}$ $\leq$ $-35.5^\circ$. In this work, a quasioptical system operating at 40--60~GHz was designed while satisfying physical constraints. The present design targets mainly ion-scale turbulence. To measure the higher end of electron-scale turbulence and fully understand cross-scale interactions, further optimisation of the quasioptical system is required.

Since EXL-50U has a large magnetic pitch angle, the mismatch angle can be large if launch angles are not carefully chosen. In this study, we found that the beam's mismatch attenuation varies with toroidal launch angle and launch frequency. This allows us to determine the optimal toroidal launch angle for a given launch frequency and density profile. We also find that to achieve matching at the core and edge with the same fixed poloidal launch angle, toroidal steering is crucial. Therefore, toroidal steering with a tunable frequency channel should be used during actual operation to minimise the mismatch angle and the mismatch attenuation. Finally, we have shown that the spatial resolution of high-$k$ measurements is adequate for this quasioptical system.
\section{Acknowledgements}
This research was supported by A*STAR through the FEAT SRTT and the SERC Central Research Fund, and was partially funded by the ENN group. Ying Hao Matthew Liang was funded by a National Science Scholarship from A*STAR, Singapore. The material is also based upon work supported by the U.S. Department of Energy, Office of Science, Office of Fusion Energy Sciences, under Award DE-SC0021150. This work is also supported by the High-End Talents Program of Hebei Province , Innovative Approaches towards Development of Carbon Free Clean Fusion Energy (No. 2021HBQZYCSB006).

\bibliographystyle{elsarticle-num} 
\bibliography{references}

\end{document}